\documentclass{scrartcl}
\usepackage[utf8]{inputenc}
\usepackage{authblk}
\usepackage[numbers,sort&compress]{natbib}
\usepackage{hyperref}
\hypersetup{colorlinks,linkcolor=[rgb]{0,0,0.5},citecolor=[rgb]{0,0,0.5},urlcolor=[rgb]{0,0,0.5},breaklinks}
\usepackage{doi}
\usepackage{amsmath}
\usepackage{graphicx}

\title{First predictive simulations for deuterium shattered pellet injection in ASDEX Upgrade}

\author[1]{M~Hoelzl}
\author[2]{D~Hu}
\author[3]{E~Nardon}
\author[3,4]{GTA~Huijsmans}
\author[6]{the JOREK Team}
\author[1,5]{the ASDEX Upgrade Team}

\affil[1]{Max Planck Institute for Plasma Physics, Boltzmannstr. 2, 85748 Garching b. M., Germany}
\affil[2]{School of Physics, Beihang University, Beijing, 100191, China}
\affil[3]{CEA, IRFM, 13108 Saint-Paul-Lez-Durance, France}
\affil[4]{Eindhoven University of Technology, P.O. Box 513, 5600 MB Eindhoven, The Netherlands}
\affil[5]{See the author list of H. Meyer et al. 2019 Nucl. Fusion 59 112014}
\affil[6]{See https://www.jorek.eu for a current list of team members}

\makeatletter
\renewcommand\maketitle{
   \begin{center}
     {\huge\sffamily\bfseries\@title\par\vspace{0.3em}}
     {\scshape\large\@author}
   \end{center}
}
\makeatother

\newcommand{\modifB}[1]{#1}
\newcommand{\modif}[1]{#1}

\begin{document}

\maketitle

\section*{Abstract}

% single paragraph, around 250 words, no math
First simulations of deuterium shattered pellet injection (SPI) into an ASDEX Upgrade H-Mode plasma with the JOREK MHD code are presented. \modifB{Resistivity is increased by one order of magnitude in most simulations to reduce computational costs and allow for extensive parameter scans}. The effect of various physical parameters onto MHD activity and thermal quench (TQ) dynamics is studied and the influence of MHD onto ablation is shown. TQs are obtained quickly after injection in most simulations with a typical duration of 100 microseconds, which slows down at lower resistivity. Although the n=1 magnetic perturbation dominates in the simulations, toroidal harmonics up to n=10 contribute to stochastization and stochastic transport in the plasma core. The post-TQ density profile remains hollow for a few hundred microseconds. However, when flux surfaces re-form around the magnetic axis, the density has become monotonic again suggesting a beneficial behaviour for runaway electron avoidance/mitigation. With $10^{21}$ atoms injected, the TQ is typically incomplete and triggered when the shards reach the q=2 rational surface. At a larger number of injected atoms, the TQ can set in even before the shards reach this surface. For low field side injection considered here, repeated formation of outward convection cells is observed in the ablation region reducing material assimilation. Responsible is a sudden rise of pressure in the high density cloud when the stochastic region expands further releasing heat from the hot core. After the TQ, strong sheared poloidal rotation is created by Maxwell stress, which contributes to re-formation of flux surfaces.

\tableofcontents

%**************************************
\section{Introduction}\label{:intro}
%**************************************

Shattered pellet injection (SPI) is the basis of the planned ITER disruption mitigation system~\cite{Lehnen2015}. Various experimental studies have been carried out for SPI in the last decade in particular at DIII-D~\cite{Commaux2010,Commaux2016,Shiraki2016} and further tokamak devices are being equipped with SPI systems presently -- an SPI system is also planned for the ASDEX Upgrade tokamak studied in this article. Nevertheless, a lot of uncertainty remains about how such a system needs to be designed for ITER in order to mitigate all negative consequences of a disruption efficiently at the same time. In particular, eddy and halo current forces need to be kept within the limits of the supporting structures, more than 90\% of the thermal energy needs to be radiated -- also a large fraction of the magnetic energy converted to thermal energy during the current quench --, and the formation of a runaway electron beam needs to be avoided or mitigated if avoidance is not possible. To match these requirements, among other constraints, the current quench (CQ) time will have to stay between given upper and lower limits and the plasma density needs to increase by more than an order of magnitude across the whole plasma domain before intact flux surfaces begin to form again when runaway electrons are confined in the plasma.

Besides NIMROD~\cite{KimC2019} and M3D-C1~\cite{Ferraro2018}, the JOREK code used for the present study is among the few extended non-linear MHD codes world-wide, which has the necessary ingredients for simulating all aspects of mitigated disruptions in realistic geometry. Previous studies for massive gas and shattered pellet injection~\cite{Reux38thEPS,Fil2015,FilPhD,Nardon2017,Hu2018,HuEPS2018} demonstrate some of the capabilities. The code allows also to study the dynamics of runaway electrons~\cite{Sommariva2017,Bandaru2019A} and vertical displacement events~\cite{ArtolaPhD,Krebs2019} which are closely related to (mitigated) disruptions. In the present article, only a small subset of the code features is invoked to perform first predictive simulations for deuterium SPI into an ASDEX Upgrade H-mode plasma. While impurity SPI will play a crucial role for ITER, also deuterium SPI is expected to be directly relevant for disruption mitigation. Such injections might allow to increase the plasma density and decrease the plasma temperature already before a thermal quench is triggered. This could be a way for mitigating both the runaway electron hot tail primary source and avalanche secondary source.

Clearly, a few aspects of the simulations will need to be further refined in the future, in particular when a lab characterization of the shard cloud becomes available for the ASDEX Upgrade injector. However the present simulations already provide considerable insights into the expected dynamics and the features of the MHD activity. In the present study, various simulations are performed and analyzed to shed light onto the dependencies of the dynamics on both plasma and injection parameters. In Section~\ref{:setup}, we describe the simulation setup used for our studies including a discussion of presently used simplifications. Section~\ref{:results} analyzes the various simulations and highlights key observations. Finally, Section~\ref{:conclusions} gives a physical interpretation of the results, and provides an outlook to more advanced simulations planned in the future.

%**************************************
\section{Simulation Setup}\label{:setup}
%**************************************

The simulations shown in this article are performed with the JOREK. non-linear extended MHD code~\cite{Huysmans2007,Czarny2008,Hoelzl2020B} using a reduced MHD model with two-fluid extensions~\cite{Orain2013} an extension for neutral particles~\cite{Fil2015} and the model for shattered pellet injection and ablation~\cite{Hu2018}. Code version 2.19.05.0 is used along with the modifications for the shattered pellet injection and ablation model described in Ref.~\cite{Hu2018}.  \modif{The JOREK reduced MHD model is based on an ansatz approach rather than an ordering approach, which generally has very good energy and momentum conservation properties (Refs.~\cite{Franck2015,Nikulsin2019}). Various benchmarks performed over the years have shown that the reduced MHD model is capturing the dynamics of MHD modes very well except for some limits in spherical tokamaks. In the context of a complete overview of the JOREK code, details about the capabilities of the JOREK reduced MHD model will be provided soon (Ref.~\cite{Hoelzl2020B}). We do not expect a significant influence of the reduced MHD model onto the results shown in the present article.} The following scalar variables are evolved in time: Poloidal magnetic flux $\Psi$, electric potential $u$, toroidal current density $j$, toroidal vorticity $\omega$, plasma density $\rho$, temperature $T$, parallel velocity $v_{||}$, and neutrals density $\rho_n$.

\begin{figure}
\centering
  \includegraphics[width=0.6\textwidth]{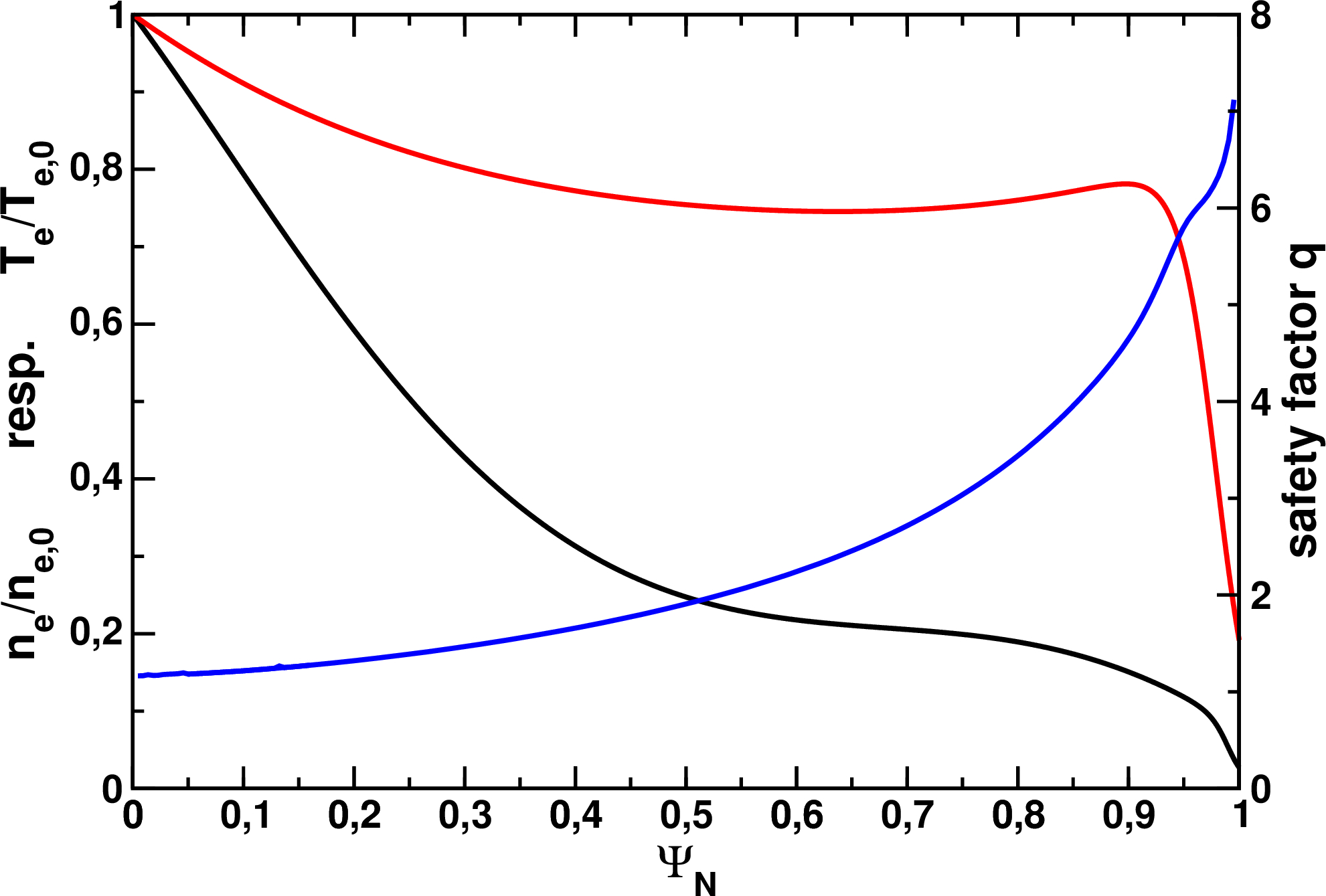}
  %\includegraphics[width=0.6\textwidth]{q-rho-T.png}
  % created with grace: ~/papers/2019-08-AUG-SPI-Paper/q-rho-T.agr
\caption{\modif{Equilibrium profiles are plotted against the normalized poloidal flux. The temperature is shown in black and the density in red. Both are normalized to their core values, see the left y-axis. The profile of the safety factor is shown in blue, see the right y-axis.}}
\label{fig:q-rho-T}
\end{figure}

A typical ASDEX Upgrade H-mode plasma is studied with a toroidal magnetic field amplitude of $2.5\mathrm{T}$, a toroidal plasma current of $0.8\mathrm{MA}$, a safety factor $q$ of about $1.2$ at the magnetic axis and about $5.8$ at $\Psi_N=0.95$, where $\Psi_N=(\Psi-\Psi_\text{axis})/(\Psi_\text{separatrix}-\Psi_\text{axis}$) denotes the normalized poloidal flux. The initial core temperature is about $3\,\mathrm{keV}$, electron and ion temperatures are assumed to be equal for simplicity, and the core electron density in the deuterium plasma is about $8\cdot10^{19}\,\mathrm{m}^{-3}$. \modif{Profiles for density, temperature, and the safety factor are shown in Figure~\ref{fig:q-rho-T}.} The injection of shattered Deuterium pellets from the low field side (LFS) mid-plane into the plasma at a vertical location of $Z=0.05$ is simulated, which is equivalent to the vertical position of the magnetic axis.

Simulations are carried out with a Spitzer-like temperature dependent resistivity $\eta=\eta_0\,(T/T_0)^{-3/2}$, where $T_0$ denotes the initial plasma core temperature.
Since the goal of this study is to perform a comparably large number of simulations and investigate the dependency of the plasma behaviour onto various SPI parameters, most simulations are performed at a resistivity of $\eta_0=2\cdot10^{-7}\,\Omega\mathrm{m}$ to limit computational costs. A single simulation is performed with a resistivity lower by a factor of ten, which approximately matches the experimental value (when taking into account neoclassical corrections; the resistivity is presently a function of the temperature only such that the radial dependency of the neoclassical corrections is not accounted for). \modif{Due to the relatively short time scales considered, perpendicular diffusion has no significant influence onto the simulation results. Perpendicular particle and heat diffusion coefficients are prescribed as functions of the normalized poloidal flux and were chosen to keep density and temperature approximately constant in time together with Gaussian source profiles. Values are in the range of $0.5\,\mathrm{m^2/s}$ for the heat diffusivity and $3\,\mathrm{m^2/s}$ for the particle diffusivity. Both values are reduced in the pedestal region}. The dynamic viscosity is given by $\mu_\text{dyn}=\mu_\text{dyn,0}\cdot(T/T_0)^{-3/2}$ where $\mu_\text{dyn,0}=4\cdot10^{-7}\,\mathrm{kg\,m^{-1}\,s^{-1}}$ in most simulations. In the simulation with lower resistivity, also the viscosity is scaled accordingly to keep the Prandtl number unchanged.

We concentrate on the following main questions \modif{and will refer back to them in the conclusions section with the attempt to provide answers}:
\begin{itemize}
    \item \modif{How does the MHD activity affect shard ablation and core plasma density increase?}
    \item \modif{Which characteristic features are observed during the TQ?}
    \item \modif{How do plasma and SPI parameters affect TQ dynamics?}
    \item \modif{At which amount of injected material is a TQ expected in the experiment?}
    \item \modif{How does the density profile evolve during and after the TQ?}
    \item \modif{How much does the core density increase before flux surface re-formation?}
\end{itemize}

Since most of these questions are related to the early evolution of the plasma before and during the early TQ, which should be dominated by dilution and stochastic losses in case of pure deuterium injection~\cite{Nardon2017}, background impurity radiation is not considered. For this reason, the post-TQ temperatures will not be fully realistic in all simulations, but the values still allow to distinguish between full and incomplete TQ. Since the focus of the work is on the pre-TQ and TQ, vertical stability is not of interest here, such that all simulations are performed in a fixed boundary setup.

The shard cloud is simplified and additional effort in matching experimental properties will be carried out in the future, when a lab characterization of the shard cloud becomes available for the planned ASDEX Upgrade injector. \modif{For simplicity, all shards in a simulation are assumed to have the same size} (see below for the number of atoms injected and the number of shards simulated). The shards are assumed to move straight with a constant velocity. The individual shards are modelled as neutrals sources with Gaussian shape (in R, Z, and $\phi$ directions) moving with a velocity $v_\text{SPI}$ into $-\mathbf{e}_R$ direction, which is $250\,m/s$ in most simulations, plus a random velocity spread of $80m/s$ added. In simulations with a single \modif{shard, i.e., an unshattered pellet, no random velocity spread is applied. The random value generator is using a user-provided ``seed'', such that simulations can be exactly reproduced if necessary. Changing the seed at otherwise identical simulation parameters allows to obtain statistical information. Based on simple separate tests, the effect introduced by this random spread is expected to be small unless a very low number of shards is injected. The shattering location is assumed to be at $R=2.155\,m$ at the $Z$ location of the magnetic axis (the shattering location in the actual AUG experiments is not yet known, but will likely be above the outboard midplane) and the injection is performed in $-\mathbf{e}_R$ direction. The shard cloud reaches the axis position after about $1.8\,\mathrm{ms}$.} The ablation model from Ref.~\cite{Gal2008} is used\modif{, where the number of atoms ablated per second for each shard is given by $4.12\cdot10^{16}\,(r_s[m])^{1.33}\,(n_e[m^-3])^{0.33}\,T_e^{1.64}$. Here, $r_s$ denotes the shard radius. A density of $6\cdot10^{28}\,\mathrm{m}^{-3}$ and a spherical shape is assumed. Further ablation models are available in the code. The ablated atoms of each shard enter the simulation via a source of neutral particles. The further ionization and recombination of these neutrals is implemented self-consistently using ADAS coefficients. Molecular processes are neglected since they are not affecting the overall energy balance significantly.}
The injection trajectory passes through the magnetic axis. The size of the Gaussian source in the poloidal plane is 8 cm and the localization in toroidal direction is 30 degrees. While the increased size of the shards in the simulations has an influence onto the simulations, also experimentally the ablation cloud is larger than the shards themselves. Furthermore, deuterium SPI is known to contain a large fraction of gas and particularly small shards, which increases the effective size of the ablation cloud.
%Additionally, simulations for the injection of single pellets~\cite{Futatani2014} have shown, that the plasma response converges for pellet sizes considerably larger than the actual physical pellet size due to the fast propagation of density and temperature perturbations in the plasma.
Finally, also turbulent transport, which may be driven by the locally enhanced gradients and might only partially be captured by the MHD model will reduce the localization of the \modif{perturbation around the shards}. Nevertheless, a more accurate modelling of the shard cloud is envisaged for future ASDEX Upgrade studies, once its experimental properties are known. 

The parallel heat diffusion coefficient $\chi_{||}$, is modelled with Spitzer-H\"arm~\cite{Spitzer1953} temperature dependency as $\chi_{||,0}\;(T/T_0)^{5/2}$. Here, the coefficient $\chi_{||,0}$ is taken to be $1\cdot10^{11}m^2/s$ in most simulations (realistic value for Spitzer-H\"arm conduction), but $\chi_{||}$ is always restricted to a maximum value of $0.1\chi_{||,0}$. \modif{This corresponds to using a maximum value for the temperature $T_\text{max}$ equal to 40\% of the initial core temperature, above which the temperature dependency of $\chi_{||}$ is dropped. The same $T_\text{max}$ value is used in all simulations}. Consequently, $\chi_{||}$ becomes temperature independent at temperatures above this threshold, while the temperature dependency in colder regions is fully retained in the simulations. This choice is taken both for numerical reasons and to avoid an over-estimation of the parallel conductivity in the hot core, i.e., it is a strongly simplified way of accounting for the so-called heat flux limit~\cite{Malone1975}. Background flows of the initial plasma are not considered at present, since a strong braking of the plasma rotation is expected anyways due to a ``locking'' of the island O-points to the injection location, which has been reported both theoretically and experimentally. Separate tests in a rotating ASDEX Upgrade X-point plasma with massive gas injection, which are not shown in this article, have confirmed, that the rotation can have a stabilizing impact if the amount of material is marginal for triggering a TQ in the absence of rotation. Under such conditions, the magnetic islands may not fully lock to the perturbation by the material injection and their rotation velocity may undergo cycles of deceleration while growing and acceleration while shrinking. When the amount of injected material is higher, the influence of the initial background rotation drops. For future simulations aiming at direct comparisons to experiments, also the effect of the background rotation will be evaluated.

To gain insights regarding the TQ triggering by the injected material, the following parameters are varied in our simulations to investigate trends:
\begin{itemize}
\item While most simulations are carried out at $\eta_0=2\cdot10^{-7}\,\Omega\,\mathrm{m}$ for computational reasons, a single simulation is performed at $\eta_0=2\cdot10^{-8}\,\Omega\,\mathrm{m}$, which approximately corresponds to the experimental value. In this simulation, also the viscosity was reduced by a factor of 10 to keep the Prandtl number identical across all simulations. In one additional simulation, the resistivity is calculated from the axisymmetric component of the temperature instead of the full 3D perturbed temperature.
\item The number of injected atoms $n_\text{atoms}$ is varied between the values $10^{21}$, $3\cdot10^{21}$, and $10^{22}$. If we assume, very roughly, that half of the material would be lost in the experiments before reaching the plasma, these parameters correspond to an initial pellet radius \modif{before shattering} of about $2\dots4.5\,\mathrm{mm}$ (assuming a spherical shape for simplicity and a density of $5\cdot10^{28}\,m^{-3}$). The smallest number of atoms studied in this article is about five times larger than that of the largest ELM pacing pellets available in ASDEX Upgrade. ELM pacing pellets in ASDEX Upgrade have $1.5\dots3.7\cdot10^{20}$ atoms~\cite{Lang2014}. Also here, the amount of material actually reaching the plasma after the $17\,\mathrm{m}$ long guide tube is reduced considerably compared to this initial particle content. The plasma particle content before the injection is about $9.6\cdot10^{20}$ atoms, the plasma volume is about $12.5\,\mathrm{m^3}$.
\item The number of shards $n_\text{shards}$ is varied in the simulations between the values $1,3,10,30,100$. Thus, also the limit of a \modif{single shard (intact pellet)} is covered.
\item The material is injected into the plasma with a velocity $v_\text{SPI}=250\,\mathrm{m/s}$ in most cases. Two simulations are carried out at $150\,\mathrm{m/s}$ and $400\,\mathrm{m/s}$ for comparison. In all cases, the shards have a velocity spread of $80\,\mathrm{m/s}$ around the average velocity (except for the cases with only a single shard, where only the nominal injection velocity is taken).
\item The toroidal field amplitude $B_\phi=F_0/R$ is varied by the choice of the parameter $F_0$, which has a value of $4.23\,\mathrm{T m}$ in most cases, corresponding to $2.48\,\mathrm{T}$ on the magnetic axis. Two simulations are carried out at $2.34\,\mathrm{T}$ and $2.58\,\mathrm{T}$, respectively, as a simple way of shifting the location of the $q=2$ rational surface. \modif{The q-value on axis is about 1.09, 1.16, and 1.21 in order of increasing magnetic field amplitude.} Thus, in all simulations, the initial safety factor on axis remains above unity to avoid the influence of an intrinsically unstable internal kink mode onto the simulations.
\item The coefficient $\chi_{||,0}$ for the parallel heat transport is usually set to $1\cdot10^{11}\,m^2/s$ in the simulations (see above for details on the model), however two simulations are carried out at 50\% and 25\% of that value, for comparison.
\item The number of Bezier finite elements $n_\text{elems}$ is about 9k in most simulations, but has been increased to about 15k in some cases to confirm convergence.
\item The toroidal Fourier harmonics included in the simulations typically cover $n=0\dots10$, but some simulations have been performed with $n=0\dots5$ and $0\dots15$ to confirm convergence.
\item Furthermore, a few simulations were carried out with an artificial current source. In these cases, the resistivity acts only on $j-j_0$ instead of acting onto the total current $j$, where $j_0$ denotes the initial current profile. Including this current source term allows to keep the background current profile nearly constant. Comparing cases with and without this current source term therefore reveals the influence of the background current profile change onto MHD mode destabilization.
\item An additional scan was performed modifying the injection geometry: case O (see Table~\ref{tab:cases}) was repeated with an injection 5, 10, 15, and 20 cm above the magnetic axis location to study the sensitivity of the MHD response with respect to off-axis injection. The results in terms of ablation, magnetic energy perturbations, and core temperature evolution are very similar to the base case with the injection directly towards the magnetic axis. For this reason, the results of this scan are only mentioned here and the data is not included in the publication.
\item To evaluate the impact of MHD activity onto the shard ablation, additional simplified simulations were carried out. In the first one, the electric and magnetic fields are fixed in time such that MHD modes cannot develop and stochastic losses of thermal energy are excluded ($\Psi$, $u$, $j$ are kept fixed). Consequently, dilution is the only relevant mechanism changing the temperature. In the second case, also the temperature (variable $T$) is kept fixed in time to investigate how ablation is affected by the change of the temperature due to dilution, parallel conduction, and MHD activity. In the third simulation, the complete background plasma is kept fixed in time during the ablation to mimic simplistic models sometimes used for ablation estimates.
\end{itemize}

Table~\ref{tab:cases} lists the key parameters of all simulations that are analyzed in Section~\ref{:results}. A few further simulations were performed in the preparation of this study, but are not shown here. In particular simulations with $1\cdot10^{22}$ atoms injected are sometimes hard to continue across the TQ due to numerical convergence issues, possibly even higher grid resolutions might be required for such simulations in the future.

% -----------------
\begin{table}
\centering
 \begin{tabular}{l|ccccccccl} 
  & $n_\text{at}$ & $n_\text{sh}$ & $\eta_0$ & $v_\text{in}$ & $\chi_{||,0}$ & $B_\phi$ & $n_\text{el}$ & $N_\text{t}$ & Remark \\
 \hline
 % A run1d21_ntor21_tgnum_zkprofneg_visco1e-6_centerchange_spivelocity_highres_lesshards
 A & .1 & 30 & 2 & .25 & 1 & 2.48 & 9 & 10 & current source \\
 A0 & .1 & 30 & 2 & .25 & 1 & 2.48 & 9 & 10 & axisym.\ resistivity \\
 % B run1d21_ntor21_tgnum_zkprofneg_visco1e-6_centerchange_v150_highres_lesshards_nocurrsrc
 B & .1 & 30 & 2 & \textbf{.15} & 1 & 2.48 & 9 & 10 & \\
 %C run1d21_ntor21_tgnum_zkprofneg_visco1e-6_centerchange_spivelocity_highres_lesshards_nocurrsrc
 C & .1 & 30 & 2 & .25 & 1 & 2.48 & 9 & 10 \\
 %D run1d21_ntor21_tgnum_zkprofneg_visco1e-6_centerchange_v400_highres_lesshards_nocurrsrc
 D & .1 & 30 & 2 & \textbf{.4} & 1 & 2.48 & 9 & 10 & \\
 \hline
 %E run3d21_n21_e1e-8_v1e-7_hr2_sp_dm2_ncs_AF_G
 E & .3 & 1 & \textbf{.2} & .25 & 1 & 2.48 & \textbf{15} & 10 & \\
 %F run3d21_n21_e1e-7_v1e-6_hr_sp_dm2_ncs_AF_G
 F & .3 & 1 & 2 & .25 & 1 & 2.48 & 9 & 10 & \\
 %G run3d21_n11_e1e-7_v1e-6_hr2_sp_dm2_ncs_AF_G
 G & .3 & 1 & 2 & .25 & 1 & 2.48 & \textbf{15} & \textbf{5} & \\
 %H run3d21_n21_e1e-7_v1e-6_hr2_sp_dm2_ncs_AF_G
 H & .3 & 1 & 2 & .25 & 1 & 2.48 & \textbf{15} & 10 & \\
 %I run3d21_n31_e1e-7_v1e-6_hr2_sp_dm2_ncs_AF_G
 I & .3 & 1 & 2 & .25 & 1 & 2.48 & \textbf{15} & \textbf{15} & \\
 \cline{3-3}
 %N run3d21_ntor21_tgnum_zkprofneg_visco1e-6_centerchange_spivelocity_highres_lessshards3_nocurrsrc
 N & .3 & 3 & 2 & .25 & 1 & 2.48 & 9 & 10 \\
 %Np run3d21_ntor21_tgnum_zkprofneg_visco1e-6_centerchange_spivelocity_highres_lessshards3_nocurrsrc_PSIFIXED_UFIXED
 Np & .3 & 3 & 2 & .25 & 1 & 2.48 & 9 & 10 & fixed vars: $\Psi$, $j$, $u$, $\omega$ \\
 %NpT run3d21_ntor21_tgnum_zkprofneg_visco1e-6_centerchange_spivelocity_highres_lessshards3_nocurrsrc_PSIFIXED_UFIXED_TFIXED
 NpT & .3 & 3 & 2 & .25 & 1 & 2.48 & 9 & 10 & fixed vars: $\Psi$, $j$, $u$, $\omega$, $T$ \\
 %NpTr run3d21_ntor21_tgnum_zkprofneg_visco1e-6_centerchange_spivelocity_highres_lessshards3_nocurrsrc_PSIFIXED_UFIXED_TFIXED_RHOFIXED
 NpTr & .3 & 3 & 2 & .25 & 1 & 2.48 & 9 & 10 & fixed vars: all \\
 \cline{3-3}
 %O run3d21_ntor21_tgnum_zkprofneg_visco1e-6_centerchange_spivelocity_highres_lessshards_nocurrsrc
 O & .3 & 30 & 2 & .25 & 1 & 2.48 & 9 & 10 \\
 \cline{3-3}
 %P run3d21_ntor21_tgnum_zkprofneg_visco1e-6_centerchange_spivelocity_highres_nocurrsrc
 P & .3 & 100 & 2 & .25 & 1 & 2.48 & 9 & 10 \\
 \hline
 % S run1d22_ntor21_tgnum_zkprofneg_visco1e-6_centerchange_spivelocity_highres_lesshards2_nocurrsrc
 S & 1 & 10 & 2 & .25 & 1 & 2.48 & 9 & 10 \\
 \cline{3-3}
 % T run1d22_ntor21_tgnum_zkprofneg_visco1e-6_centerchange_spivelocity_highres_lesshards_nocurrsrc_zkpar1e4
 T & 1 & 30 & 2 & .25 & \textbf{.25} & 2.48 & 9 & 10 \\
 % U run1d22_ntor21_tgnum_zkprofneg_visco1e-6_centerchange_spivelocity_highres_lesshards_nocurrsrc_zkpar3e4
 U & 1 & 30 & 2 & .25 & \textbf{.5} & 2.48 & 9 & 10 \\
 % V run1d22_ntor21_tgnum_zkprofneg_visco1e-6_centerchange_spivelocity_highres_lesshards_nocurrsrc
 V & 1 & 30 & 2 & .25 & 1 & 2.48 & 9 & 10 \\
 \cline{3-3}
 % W run1d22_ntor21_tgnum_zkprofneg_visco1e-6_centerchange_spivelocity_highres_nocurrsrc
 W & 1 & 100 & 2 & .25 & 1 & 2.48 & 9 & 10 \\
 % X run1d22_ntor21_tgnum_zkprofneg_visco1e-6_centerchange_spivelocity_highres_nocurrsrc_F0+4.0
 X & 1 & 100 & 2 & .25 & 1 & \textbf{2.34} & 9 & 10 \\
 % Y run1d22_ntor21_tgnum_zkprofneg_visco1e-6_centerchange_spivelocity_highres_nocurrsrc_F0+4.4
 Y & 1 & 100 & 2 & .25 & 1 & \textbf{2.58} & 9 & 10 \\
 \hline
 \end{tabular}
 \caption{An overview is given of the parameters used for all simulation cases investigated in this study. The following information is shown: $n_\text{at}$ is the number of atoms in the \modif{original pellet before shattering} in units of $10^{22}$, $n_\text{sh}$ represents the number of shards \modif{(all shards have the same size as described in the main text)}, $\eta_0$ is the initial core resistivity in units of $10^{-7}\,\Omega\mathrm{m}$, $v_\text{in}$ is the injection velocity in units of $\mathrm{km/s}$, $\chi_{||,0}$ the parallel conductivity coefficient in units of $10^{11}\,m^2/s$, $B_\phi$ denotes the toroidal field strength on axis in units of $\mathrm{T}$,  $n_\text{el}$ gives the number of grid elements in units of $1000$, $N_\text{t}$ refers to the largest toroidal mode number included in the simulation, $j_0$ indicates whether a source is applied to keep the current profile approximately constant in time. In the column ``remark'', special setups of the respective cases are mentioned, e.g., those variables are listed, which are kept constant in time. Parameter values deviating from the values used in most simulations are highlighted in bold print.}
 \label{tab:cases}
\end{table}
% -----------------

%**************************************
\section{Results}\label{:results}
%**************************************

In this Section, the simulation results are analyzed. Subsection~\ref{:results:converg} discusses numerical convergence of the simulations. Subsection~\ref{:results:abl} investigates the impact of MHD activity onto the \modif{pellet/shard} ablation. Subsection~\ref{:results:dens} analyzes the evolution of the plasma particle content across all simulations. Subsection~\ref{:results:tq} studies key properties of the thermal quench observed in the simulations. Subsection~\ref{:results:averaged} analyzes the evolution of poloidally and toroidally averaged profiles. And finally, Subsection~\ref{:results:dynamics} describes the detailed dynamics of a selected case that is associated to a strong increase of the plasma particle content.

%**************************************
\subsection{Convergence}\label{:results:converg}
%**************************************

\begin{figure}
\centering
  \includegraphics[width=0.6\textwidth]{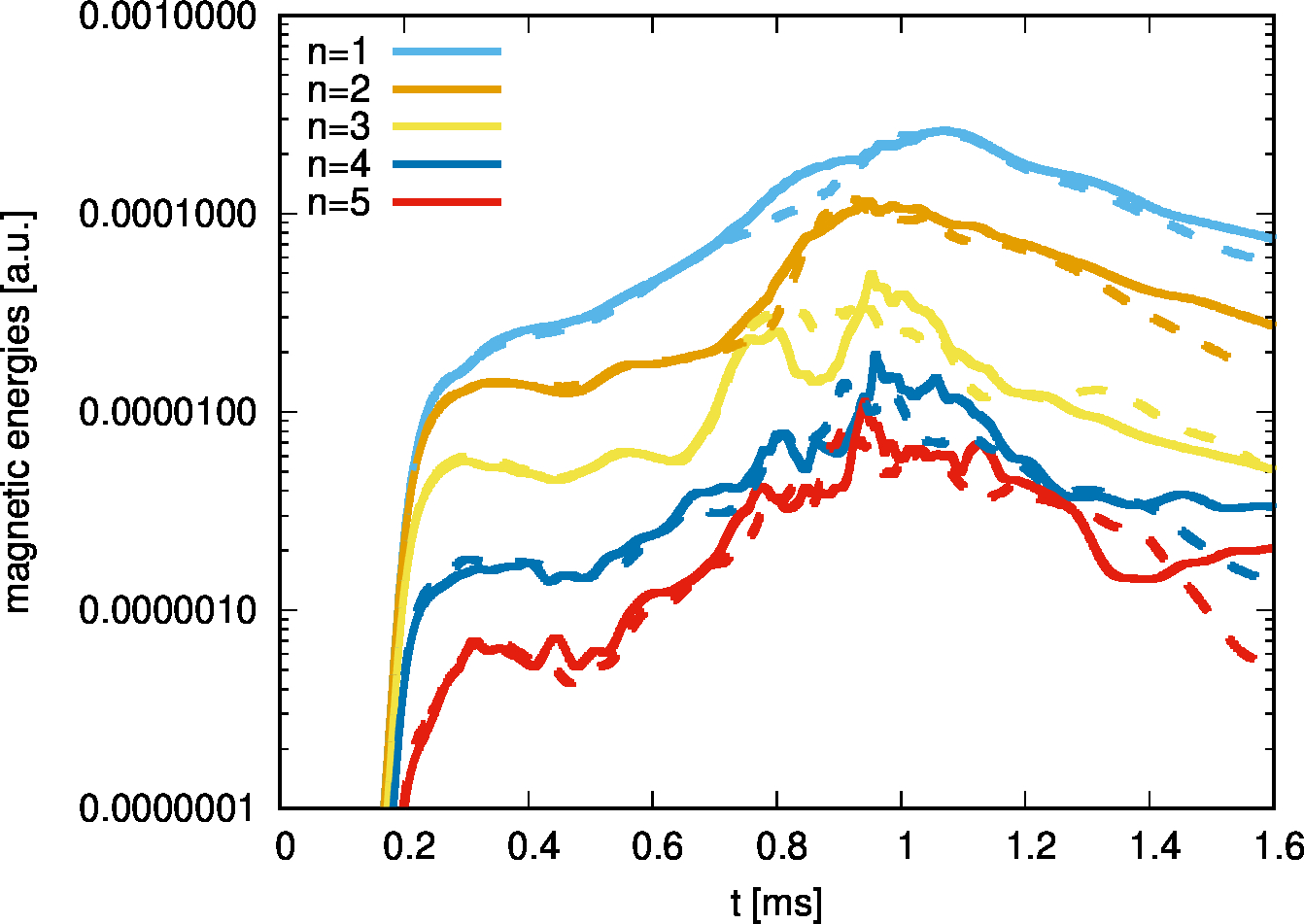}
  %\includegraphics[width=0.5\textwidth]{F-I-energies.jpg}
  % created with gnuplot by 2019-01-29-AUG-SPI/plot-F-I-energies.gp
\caption{Comparison of the $n=1\dots5$ magnetic energies between cases F and I. The color coding indicates the mode number. Solid lines correspond to case F and dashed lines to case I. Both simulations show very good agreement in spite of the different poloidal and toroidal resolutions.}
\label{fig:F-I-energies}
\end{figure}

Before analyzing the results of the simulations, a few basic tests are performed to confirm numerical convergence. Poloidal and toroidal resolutions and the size of the time step were varied to confirm that the results are converged very well in the perturbed magnetic energies during the non-linear evolution. Hyper-resistivity and hyper-viscosity are included for numerical reasons only with very small coefficients and scans in both parameters confirmed that the physical results are not affected by these parameters. For cases with $\eta=2\cdot10^{-7}\,\Omega\,\mathrm{m}$, a resolution of $n=0\dots10$ and 9k grid elements is sufficient. As an example, the $n=0\dots5$ magnetic energies of a case with 9k grid elements and the harmonics $n=0\dots5$ included in the simulation are compared to a case with 15k grid elements and $n=0\dots15$ included in the simulation. As visible from Figure~\ref{fig:F-I-energies}, very good agreement is observed before, during, and after the TQ \modif{(note, however, that the TQ duration is shorter in the simulation with a higher resolution)}. In production simulations, nevertheless, a toroidal resolution of $n=0\dots10$ is used in the following, since toroidal mode numbers up to 10 may still contribute considerably to the stochastic radial transport although they are sub-dominant. For cases with a resistivity of $2\cdot10^{-8}\,\Omega\mathrm{m}$, 15k grid elements are required for numerical stability reasons, while the toroidal resolution requirement does not change.

%Cases F-I are all performed at identical physical parameters. Case H is performed with 15k grid elements and a toroidal resolution of $n=0\dots10$. Case F has a reduced poloidal resolution of 9k node elements. Case G has a reduced toroidal resolution of $n=0\dots5$, and case I has an increased toroidal resolution of $n=0\dots15$. In all cases, the $n=1$ magnetic energy is the dominant component throughout the simulation.
%
%The $n=1$ magnetic energy and also the sum of all non-axisymmetric magnetic energies is very similar (deviation of a few percent) across all four cases in the early phase, when the ablation starts in the edge of the plasma and the edge plasma confinement is lost due to stochastization. In the further evolution of the pre-TQ phase, case G shows a stronger magnetic perturbation than the cases F, H, and I, which remain very close together (again in the range of a few percent difference). This confirms, that the results are converged at a poloidal resolution of 9k grid elements, and a toroidal resolution $n=0\dots10$ in this case during the pre-TQ phase.
%
%During the TQ ...
%
%Separate tests have shown, that 15k grid elements are needed at a the cases with a ten times lower resistivity.
%
%Cases J and K are identical except for the numerical time stepping scheme used (Crank-Nicholson in the first and Gears in the second case). The deviations between both cases are negligible.

Furthermore, a few tests were carried out to verify the dependence of the results onto the plasma resistivity. Due to the computational costs, only few of those cases were run into the TQ at fully realistic resistivities. The tests show that results are close together early in the simulation when ablation starts. However at the lower resistivity, the subsequent rise of the magnetic perturbation takes place slower and the TQ is delayed. Also, the amount of material required for triggering a \modif{(prompt)} TQ is higher at lower resistivity. Tests confirm, that the plasma dynamics at increased resistivity values are similar to the dynamics at realistic values, however a lower number of atoms is sufficient for causing an MHD response of comparable amplitude. The available data indicates that about two to three times more material is needed for triggering a \modif{(prompt)} TQ at fully realistic resistivity compared to the ten times higher resistivity used in most simulations. This corresponds to an increase of the (spherical) pellet radius by $20\dots40$\%. Further simulations at fully realistic parameters will be carried out in the future, when the specification of the ASDEX Upgrade injector are fully known and presently ongoing optimizations of the solver have increased the computational efficiency for simulations at high resolution.

%e.g., E, H -- different eta; TQ no TQ!

%**************************************
\subsection{Ablation with a changing background}\label{:results:abl}
%**************************************

To investigate how the change of the temperature (and density) during injection affects the ablation of the shards, several simulations were performed with the same equilibrium and shard cloud parameters. Besides a fully consistent MHD simulation (case N; see parameters in Table~\ref{tab:cases}), three additional simulations are performed, where the model was successively simplified. In case Np, the setup is identical to case N, however the electric field is zero and the magnetic field remains constant throughout the simulation. In case NpT, also the temperature is kept fixed in time. And finally in case NpTr all variables are kept constant in time such that the ablation is calculated in the absence of any perturbations of the initial plasma state.

% -----------------
\begin{table}
\centering
 \begin{tabular}{lc|c|c|c|c|c|c|c|c|c|c|} 
  & & \multicolumn{6}{c|}{Temperature} & \multicolumn{4}{c|}{Density} \\
  \hline
  & & dil. & $\chi_\bot$ & $\chi_{||}$ & $v_{\bot}$ & $v_{||}$ & $P_\text{Ohm}$ & $D$ & $v_\bot$ & $v_{||}$ & $S_{SPI}$ \\
 \hline
 N    & (MHD)         & y & y & y   & y & y   & y & y & y & y   & y \\
 Np   & ($B$ fixed)   & y & y & (y) & - & (y) & y & y & - & (y) & y \\
 NpT  & ($B,T$ fixed) & - & - & -   & - & -   & - & y & - & (y) & y \\
 NpTr & (all fixed)   &- & - & -   & - & -   & - & - & - & -   & - \\
 \end{tabular}
 \caption{The tables shows which of the relevant terms, that can affect the temperature and density distributions, are retained in the four simulations considered in Section~\ref{:results:abl}. Here ``y'' indicates that the term is retained, ``(y)'' indicates that the term is retained without reflecting stochastic transport since the magnetic field is kept fixed in time, and ``-'' indicates that the term is not present in the simulation.}
 \label{tab:Netc}
\end{table}
% -----------------

\begin{figure}
\centering
  \includegraphics[width=0.6\textwidth]{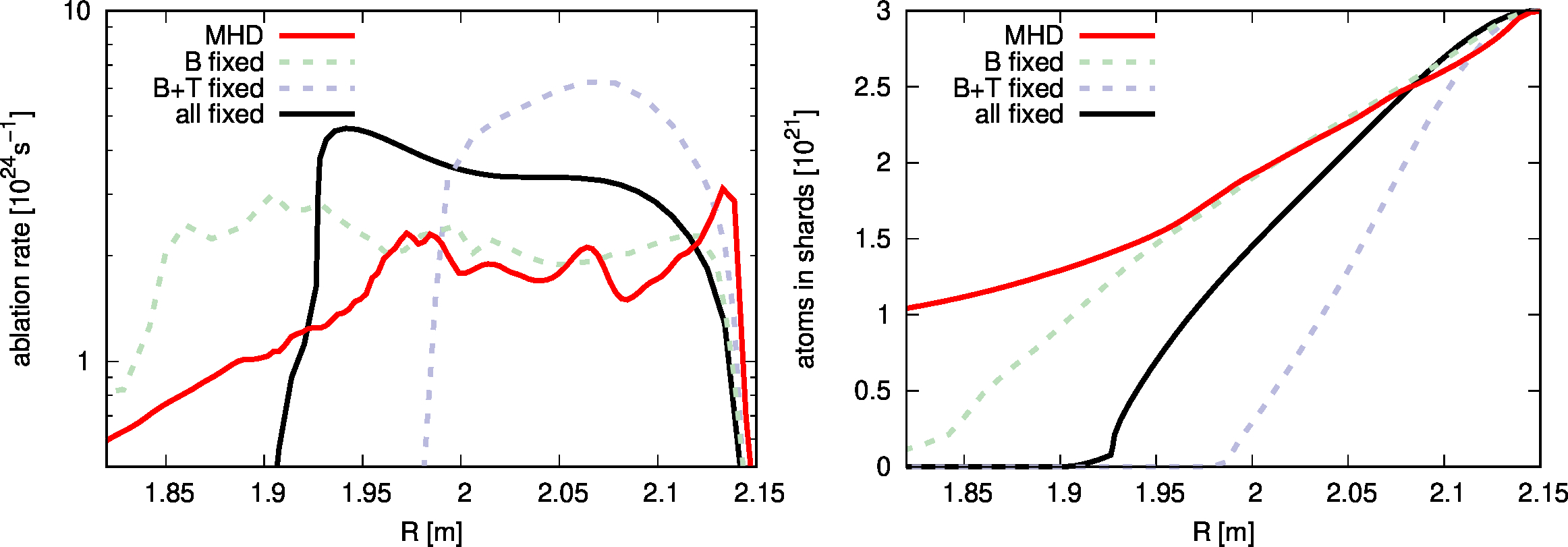}
  %\includegraphics[height=0.34\textwidth]{ablation_N_Np_NpT.jpg}
  %\vspace{1em}
  %\includegraphics[height=0.34\textwidth]{atomsleft_N_Np_NpT.jpg}
  % created with gnuplot by 2019-01-29-AUG-SPI/compare_N_Np_NpT.gp
\caption{The ablation of several simulations is compared. Case N is shown with a red solid line, case Np with a green dashed line, case NpT with a grey dashed line, and case NpTr with a black solid line.  Injection is from the right. Left: Ablation rate from the three shards injected in the cases discussed in Subsection~\ref{:results:abl}. Right: Number of atoms left in the shards for the same simulations. Clearly, ablation is only captured correctly when MHD activity is properly taken into account.}
\label{fig:Netc}
\end{figure}

The temperature in the MHD simulation can change due to the ionization energy (negligible for deuterium), radiation (negligible for deuterium), dilution, perpendicular diffusion, parallel diffusion, perpendicular convection, parallel convection and Ohmic heating. The density can change due to parallel and perpendicular convection, isotropic diffusion, ionization of neutrals \modif{(mostly neutrals resulting from the ablation of the shards)}, and recombination (negligible at the post-TQ temperatures obtained here). Table~\ref{tab:Netc} shows which of the terms are retained in the four cases considered here. From case N to case Np, perpendicular convection and stochastic transport are dropped, which does not allow for a true thermal quench to occur. From case Np to NpT, all changes of the temperature are dropped while the density can still evolve. And finally, case NpTr reflects ablation in the presence of a completely static background. Figure~\ref{fig:Netc} compares the ablation rates and shard sizes in the four cases during injection. The location of the shard cloud is shown on the X-axis, injection is from the right.

In the very beginning of the injection, the shards are located in a cold plasma region (SOL or pedestal bottom) such that ablation is slow. However, in case N, where stochastization of the plasma edge sets in quickly, heat from the pedestal region is released along the magnetic field lines into the SOL such that ablation increases in this case compared to the other simulations. Case NpT successively shows the fastest ablation, since the density can increase by the ablation (which enhances ablation in turn), but the temperature remains unperturbed. The injection with completely fixed background (case NpTr) exhibits only slightly lower ablation rates than case NpT, \modif{since the ablation rate exhibits only a weak density dependence}. Case Np initially shows very similar ablation to the fully consistent MHD simulation (case N). However, this changes when the TQ sets in: Ablation in case N drops quickly following the core TQ onset, since the temperature has dropped, while ablation in case Np remains high until complete ablation since an MHD driven thermal quench is not present in that case (temperature only drops by dilution in case Np). Due to the absence of impurity background radiation, the post-TQ temperatures are overestimated in the simulations, such that even case N still exhibits a small ablation rate after the TQ. Either way, the shards in case N are not fully ablated and the remaining material would eventually hit the plasma facing components on the opposite side of the injection location. \modif{The tests show that simplified lower-dimensional models, which do not take into account the TQ dynamics at all, or simulations with an unrealistically low parallel heat diffusion coefficient which is not capturing stochastic losses correctly, tend to strongly over-predict pellet ablation.}

%**************************************
\subsection{Particle content}\label{:results:dens}
%**************************************

\begin{figure}
\centering
  \includegraphics[width=0.6\textwidth]{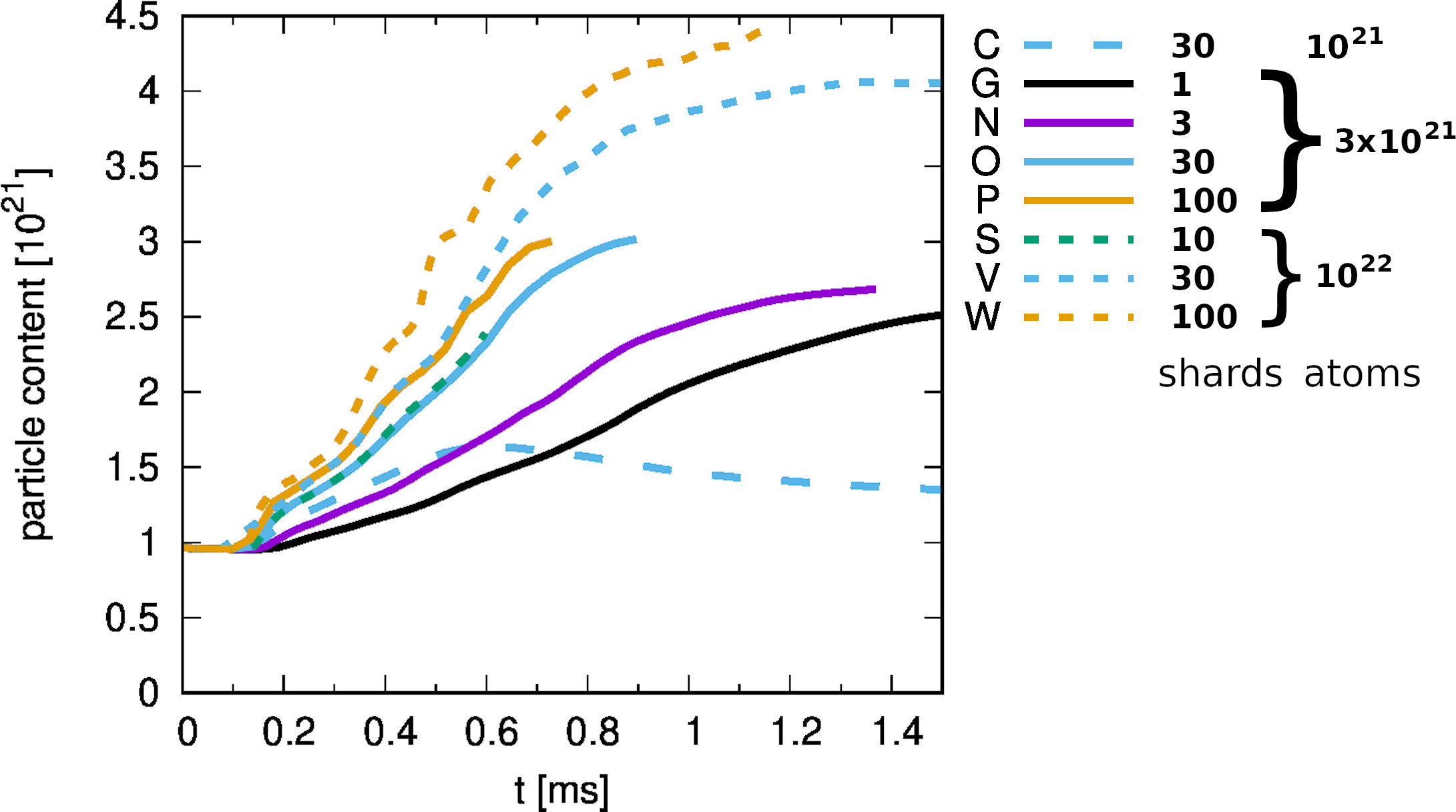}
  %\includegraphics[width=0.75\textwidth]{particlecontents_selected.jpg}
  % created with gnuplot by 2019-01-29-AUG-SPI/all-particlecontents.gp
\caption{Plasma particle content is plotted versus the simulation time for various simulations, which differ only by the amount of injected material and the number of shards (same cases as in Figure~\ref{fig:Tcorerhocore}).}
\label{fig:partcont}
\end{figure}

In this Section, the evolution of the particle content is compared across various simulations \modif{to provide a first overview of the different cases}. Figure~\ref{fig:partcont} shows the time evolution of the particle content for many different simulations over time. Clearly, the amount of injected material and the number of shards strongly influence, how much material is assimilated. \modif{With an increasing number of shards, the overall surface area increases such that ablation becomes more efficient and the rise of the particle content is accelerated. Consequently, also the perturbation amplitudes increase and the destabilization of MHD activity is accelerated. With an increasing shard number, a convergence trend of the particle content starts to become visible, i.e., the evolution of the particle content becomes weakly dependent on the number of shards.}

Case W with $1\cdot10^{22}$ atoms injected in 100 shards exhibits the strongest increase of the particle content, by a factor of almost 4.5. Most simulations are stopped right after the TQ to save computational time although they could be continued. For those cases which were continued further, finite ablation is still observed after the TQ. However, with background impurity radiation included, the post-TQ temperature would be lower in most cases such that ablation came to an end (\modif{the temperature drop would likely remain similar for case C when accounting for impurity radiation, since the incomplete TQ observed here does not lead to a strong enough temperature collapse to reach temperature values where impurity radiation peaks}).

\begin{figure}
\centering
  \includegraphics[width=0.6\textwidth]{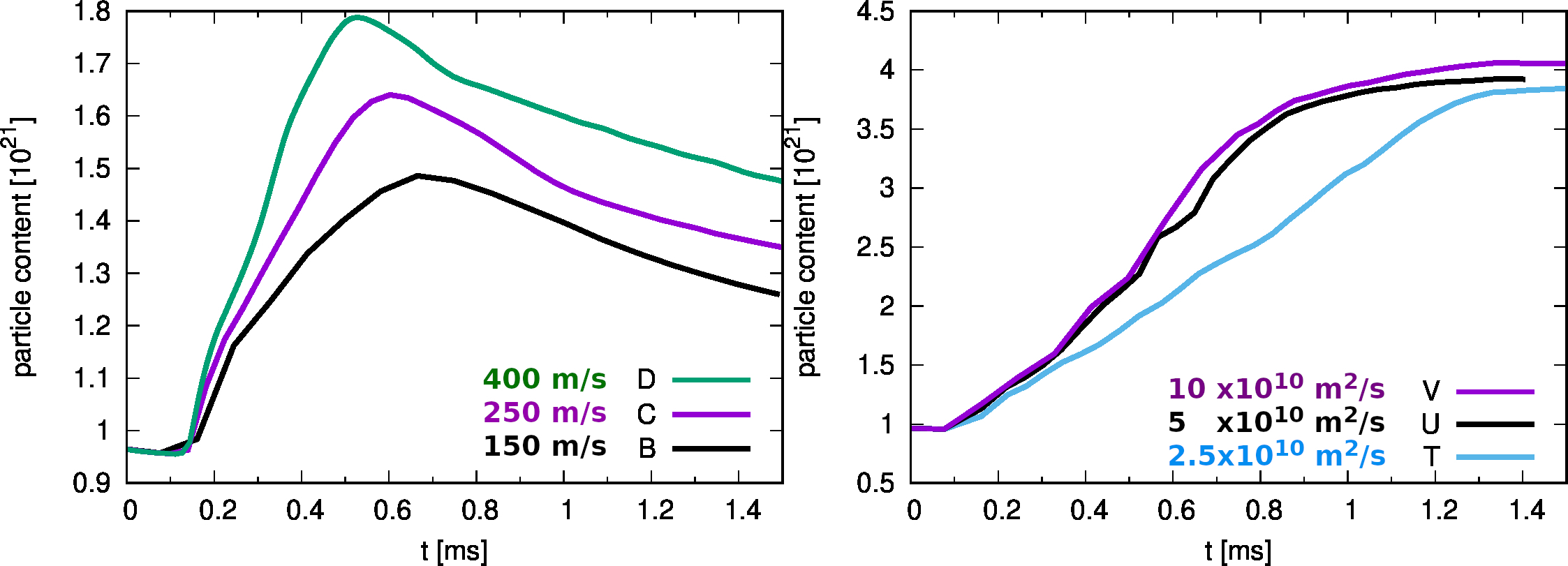}
%\includegraphics[width=0.45\textwidth]{particlecontents_resol.jpg}
%\vspace{1em}
%\includegraphics[width=0.45\textwidth]{particlecontents_vel.jpg}
%\vspace{1em}
%\includegraphics[width=0.45\textwidth]{particlecontents_eta.jpg}
%\vspace{1em}
%\includegraphics[width=0.45\textwidth]{particlecontents_chipar.jpg}
%\vspace{1em}
%\includegraphics[width=0.45\textwidth]{particlecontents_F0.jpg}
  % created with gnuplot by 2019-01-29-AUG-SPI/all-particlecontents.gp
\caption{The plasma particle content is plotted versus the simulation time for simulations which only differ in a single parameter.
%Top left: Simulations with different poloidal/toroidal resolutions. Top right: Different injection velocities. Mid left: Different resistivities. Mid right: Different parallel heat conductivities. Bottom: Different toroidal field amplitudes.
\modif{Left: Different injection velocities. Right: Different parallel heat conductivities.}}
\label{fig:partcontscans}
\end{figure}

\modif{In Figure~\ref{fig:partcontscans}, two comparisons between different cases are shown. Each plot shows a set of simulations which are identical except for a single parameter. The injection velocity strongly influences the assimilated particle content, since the shards can reach plasma regions with higher temperatures faster such that ablation is enhanced. When ablation takes place deeper inside the plasma, material assimilation is enhanced as well, since successive losses of ablated material are reduced. As a result, the maximum particle content is strongly increased for higher injection velocities. Also the parallel heat diffusion coefficient plays an important role since a larger ``reservoir'' of thermal energy is available for ablation when the transport is faster: In the limit of fast parallel heat transport, the thermal energy of a whole flux surface can contribute to shard ablation, while only a fraction of it can contribute at low parallel conductivities. In a stochastic state, higher parallel heat conductivity also increases the radial region that can contribute thermal energy for the ablation process. Temperature cannot equilibrate quickly enough along field lines in the low conductivity limit, which slows down ablation. While the difference between the simulation with fully realistic conductivity, and the simulation with half of this value is small, the simulation with the parallel heat conductivity further reduced by a factor two shows a strongly delayed ablation. At the same time, lower parallel conductivities also decrease the stochastic radial losses. Scans regarding the influence of the toroidal resolution (cases I, H, G, F), the resistivity (cases E, F), and the toroidal field strength (cases W, X, Y) show only negligible influence onto the evolution of the total particle content and are not plotted for that reason.}

%**************************************
\subsection{Thermal quench}\label{:results:tq}
%**************************************

% -----------------
\begin{table}
\centering
 \begin{tabular}{r|ccc} 
 case & $t_\text{TQ start}$ & $\Delta t_\text{TQ}$ & $T_\text{post TQ}$ \\
 \hline
  A & \multicolumn{3}{c}{(no TQ before $1.30\mathrm{ms}$)} \\
  A0 & \multicolumn{3}{c}{(no TQ before $1.30\mathrm{ms}$)} \\
  B & \multicolumn{3}{c}{(no TQ before $1.45\mathrm{ms}$)} \\
  C & 931 & 316 & 27\% \\
  D & 527 & 119 & 18\% \\
  \hline
  E & 1176 & 591 & 5\% \\
  F & 859 & 91 & 12\% \\
  G & 851 & 95 & 11\% \\
  H & 860 & 45 & 13\% \\
  I & 883 & 40 & 14\% \\
  N & 709 & 96 & 10\% \\
  O & 596 & 95 & 9\% \\
  P & 576 & 100 & 8\% \\
  \hline
  S & 594 & ? & ? \\
  T & 703 & 236 & 5\% \\
  U & 576 & 108 & 9\% \\
  V & 565 & 114 & 5\% \\
  W & 516 & 75 & 6\% \\
  X & 455 & 323 & 6\% \\
  Y & 515 & 124 & 5\% \\
 \hline
 \end{tabular}
 \caption{An overview is given of the TQ behaviour for the cases of Table~\ref{tab:cases}. For those cases which show a core TQ, the TQ onset time and the TQ duration are given in $\mu\mathrm{s}$. The TQ onset is defined as the point in time, at which 5\% of the fast core temperature drop has completed, and the TQ completion is the time when 95\% of the core temperature drop has taken place. The post TQ temperature is given as a fraction of the initial core temperature, which is around $3\,\mathrm{keV}$ in the considered equilibrium. Case S did not complete the TQ for numerical reasons.}
 \label{tab:tqtimes}
\end{table}
% -----------------

This Section compares the TQ \modif{properties} across the whole set of simulations. A first rise of magnetic energies to significant amplitudes is observed already at a simulation time of about $0.15\,\mathrm{ms}$ corresponding to the excitation of edge instabilities, which lead to a stochastization of the edge plasma and a loss of the edge plasma confinement. This happens almost immediately, when the first shards reach the plasma boundary. Thus, an ``edge TQ'' is observed very quickly after injection. Likely, this high sensitivity of the plasma to the perturbation is caused by the steep pedestal profiles of the H-mode plasma considered here. The edge crash has similarities to pellet triggered ELMs (simulations shown, e.g., in Ref.~\cite{Futatani2019}), however, the triggering is accelerated by the large amount of injected material in the present study \modif{and ablation is enhanced in addition by the larger overall surface area of the shards compared to a single pellet}. The evolution of magnetic perturbation energies and Poincar\'e plots are shown in Section~\ref{:results:dynamics}.

\begin{figure}
\centering
  \includegraphics[width=0.6\textwidth]{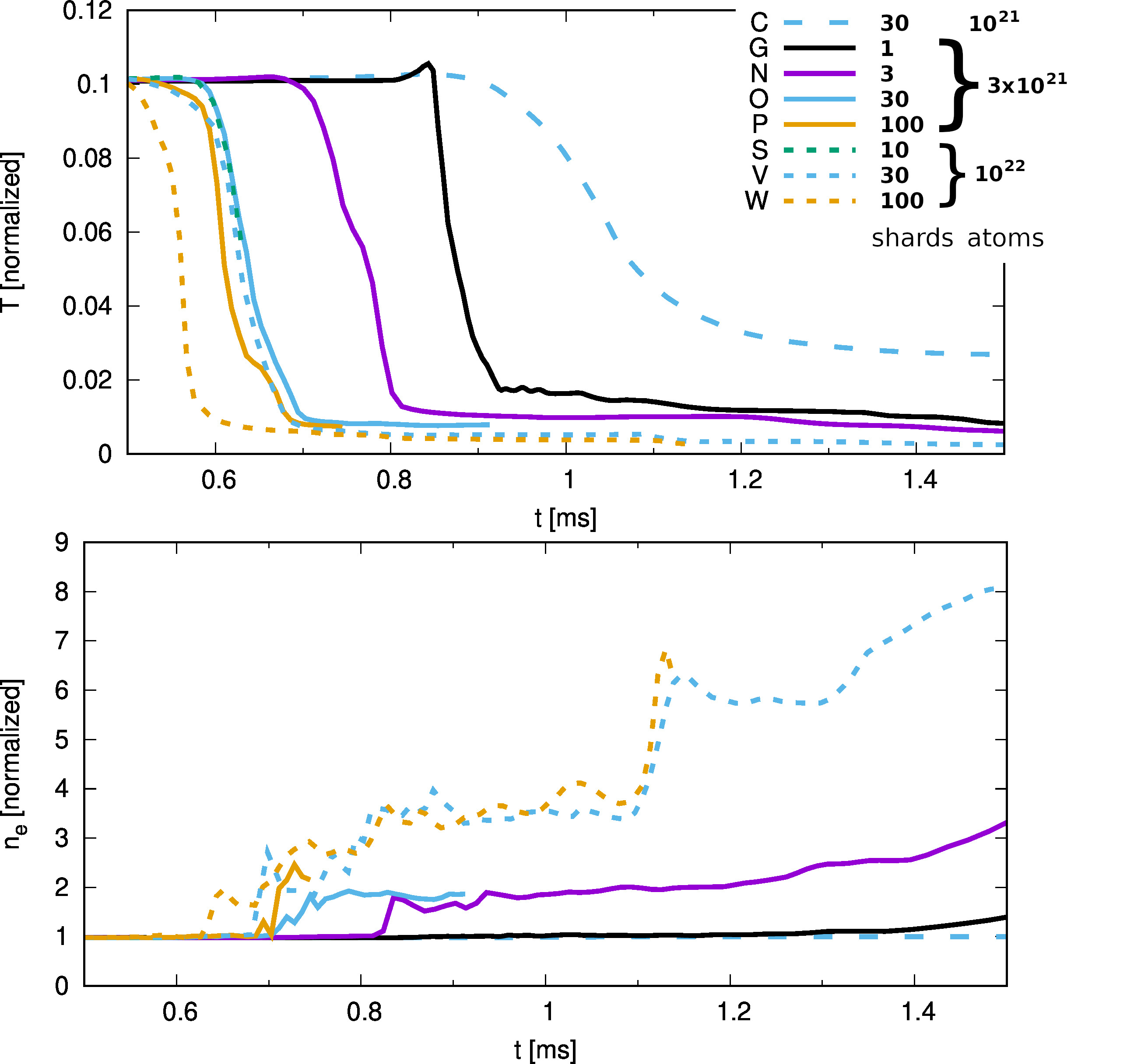}
  %\includegraphics[width=0.85\textwidth]{Tcore-selected.jpg}
  %\includegraphics[width=0.85\textwidth]{rhocore-selected.jpg}
  % created with gnuplot by 2019-01-29-AUG-SPI/all-Tcore-rhocore.gp
\caption{The core plasma temperature and density are plotted versus the simulation time for various simulations, which differ only by the amount of injected material and the number of shards. Dashed lines correspond to $10^{21}$ injected atoms, solid lines to $3\cdot10^{21}$, and dotted lines to $10^{22}$. The color coding is the following: single shard in black, three shards in magenta, ten shards in green, 30 shards in blue, and 100 shards in orange. Some simulations are stopped after completion of the TQ to save computational time such that curves do not continue further.}
\label{fig:Tcorerhocore}
\end{figure}

The ``core TQ'' sets in around $0.5\dots1\,\mathrm{ms}$ in the simulations, and completes within typically $0.1\,\mathrm{ms}$ like shown in Table~\ref{tab:tqtimes} and seen from Figure~\ref{fig:Tcorerhocore}. The magnetic perturbation energies during the TQ are higher by about one order of magnitude compared to those of the edge crash.
Close to the threshold for obtaining a TQ (e.g., by reducing the amount of injected atoms), the TQ duration increases strongly and the TQ may become incomplete with post-TQ temperatures $>10\%$ of the initial core temperature. Like mentioned before, the post-TQ temperature is not captured fully accurately in the present simulations, since radiative cooling by background impurities is not taken into account. However, it is still a good indicator showing whether stochastization is strong enough and/or long-lived enough to create a full TQ, or only a partial TQ (in case of a partial TQ, the temperatures are still so high that background impurity radiation would likely not be effective depending on the impurity species present). In this study, only a ``prompt'' core TQ shortly after injection is considered. Plasmas which are not quenching that quickly might still develop a TQ a few milliseconds later. On the other hand, a plasma undergoing an incomplete TQ (like cases C or D), might recover in a way comparable to the plasma recovery from partial thermal quenches/minor disruptions~\cite{Sweeney2018}. Such a longer term evolution of the plasma is not investigated in the present study.

%Boozer papers:
%in particular Bnonideal growing exponentially faster than current
%triggering time
%magnetic helicity conservation: volumentric term and surface term
%surface: advection, loop voltage
%j||/B constant along field lines, not any more in field line joining; leads to a Lorentz force causing Alfven waves, which are dissipated non-ideally
%energy loss by B-field due to j.E /=0 (particle acceleration)
%dK||/dt = vk K|| + q v|| E||m where E||m is far larger than the actual parallel electric field
%first term also in the ideal case possible!

In the simulations with 250 m/s injection velocity, the shards reach the rational surfaces approximately at 0.3 ms (q=4), 0.45 ms (q=3), 0.8 ms (q=2), and 1.3 ms (q=1.5). Those numbers are calculated for the q-profile of the initial equilibrium. Due to actual changes of the equilibria, these times differ slightly between cases. A detailed comparison shows that these TQ onset times correspond to the shard cloud reaching the q=2 surface in cases with a lower amount of atoms or fewer shards (cases C and F/G/H/I).

In all cases with either more atoms or more shards, the TQ takes place already earlier. In cases N and T, the shards have crossed the $q=2.5$ surface already, in cases O,P and U-Y, the shards have only reached the region of the $q=3$ or $q=2.5$ surfaces. Cases A-B do not show a TQ within the simulated time, while the shards have been ablated completely already well before the end of these simulations.

Only in the simulation with $\eta=2\cdot10^{-8}\,\Omega\mathrm{m}$ and a single shard with $3\cdot10^{21}$ atoms, the TQ onset is observed after the shard has crossed the $q=2$ surface. More atoms or shards would likely lead to an earlier TQ again.

In most cases, the core TQ completes within about $0.1\,\mathrm{ms}$ after the onset. A small amount of atoms (case C), or a low resistivity with few shards and a moderate number of atoms (case E) increase this time. In cases T and X, the TQ takes place in two ``steps'' such that the overall duration seems longer, while the first crash is still faster than $0.1\,\mathrm{ms}$. \modif{From our simulations we were not able to identify a causal link between particular parameters and a step-like sub-structure in the TQ. Such a direct link seems hard to find due to the highly non-linear nature of the problem.}

The post-TQ temperature shows a strong dependency on the number of injected atoms (and also some dependency on the number of shards). In particular the cases C and D exhibit only an incomplete TQ. At lower resistivity, the TQ onset is later and the TQ duration is considerably longer such that the post-TQ temperature is still low in this case.

%XXXXXXXXXXXXXXXXXXXXXXXXXX
%eta=1e-8 cases with 3 and 10 shards???
%XXXXXXXXXXXXXXXXXXXXXXXXXX

Cases B, C, and D allow to compare different injection velocities. The slowest injection of $150\,\mathrm{m/s}$ does not show a TQ within the simulated time frame, the injection at $250\,\mathrm{m/s}$ shows a rather late and slow TQ, and the injection at $400\,\mathrm{m/s}$ shows an earlier and faster TQ leading to a slightly lower post-TQ temperature.

The comparison between cases A and C shows the importance of changes in the current profile. Indeed, while case A (which has a current source set up to approximately maintain the initial current profile) does not show a TQ, case C (which does not have such a source term) exhibits a partial TQ when the shards reach the $q=2$ surface. Consequently, the steepening of the current profile caused by the edge cooling is an important factor in the destabilization of the MHD activity leading to the TQ. In case A0, where the helical perturbation of the temperature is not taken into account when calculating the plasma resistivity, a TQ is not observed. This confirms, that also the helical current perturbation resulting from the helical temperature perturbation (via the temperature dependent resistivity) plays a crucial role for destabilizing the MHD modes.

The effect of shifting the 2/1 rational surface can be investigated by comparing cases W, X, and Y. The results are not conclusive here, however. Possibly a scan with fewer atoms, where a stronger correlation of the TQ onset to the shards reaching the $q=2$ surface was seen, would lead to a more conclusive picture. However, since the central safety factor is changing as well in this scan (the toroidal field strength is varied, while all other aspects of the equilibrium are kept the same), the comparison must be anyway be seen as preliminary and is mentioned here only for completeness. 

To investigate the impact of the parallel conductivity onto the plasma dynamics, the three cases T ($\chi_{||,0}=2.5\cdot10^{10}\,\mathrm{m^2/s}$), U ($5\cdot10^{10}\,\mathrm{m^2/s}$), and V ($1\cdot10^{11}\,\mathrm{m^2/s}$) are considered. Ablation takes place significantly slower in case T compared to the other two cases as seen in Figure~\ref{fig:partcontscans}. As a result, T shows a delayed TQ and a slower TQ. Cases U and V show very similar ablation, TQ onset times, and TQ duration indicating that above a particular threshold, the parallel conductivity does not affect the MHD activity strongly any more. Nevertheless, the magnetic perturbation energy is higher in case V than in case U by a factor of almost two. Both, this increased perturbation amplitude (stronger stochastization) and the higher parallel conductivity act together to reduce the power-TQ temperature of case V to lower values than in case U, by almost a factor two.

\modif{In this section, the TQ dynamics of the various simulations was analyzed to draw conclusions on parameter dependencies. In general, the TQ is observed early in the simulations when the shards reach the $q=2$ surface or even earlier and the TQ takes place within typically $0.1\,\mathrm{ms}$. At lower resistivity values, the TQ onset is delayed and the TQ duration increased. At a small amount of injected material, incomplete TQs are observed. With increasing injection velocity, material assimilation is enhanced significantly and the TQ takes place earlier. Both, the change of the axisymmetric current profile and the helical current perturbation are shown to play an important role for triggering strong MHD activity and the TQ. Variations of the toroidal magnetic field amplitude to shift the radial location of the rational surfaces did not give conclusive results. Simulations with realistic parallel heat conductivities exhibit increased ablation, an earlier TQ, and lower post-TQ temperatures compared to simulations with slower parallel transport.}

Figure~\ref{fig:Tcorerhocore} shows the evolution of the plasma core temperature and densities in comparison for several cases. This will be analyzed in more depth in the following Sections. In Section~\ref{:results:averaged}, the evolution of the background profiles is analyzed for particular cases, and finally in Section~\ref{:results:dynamics}, the dynamics are investigated in more detail for a selected case associated with a strong increase of the plasma particle content.

%**************************************
\subsection{Evolution of poloidally and toroidally averaged profiles}\label{:results:averaged}
%**************************************

\begin{figure}
\centering
  \includegraphics[width=0.6\textwidth]{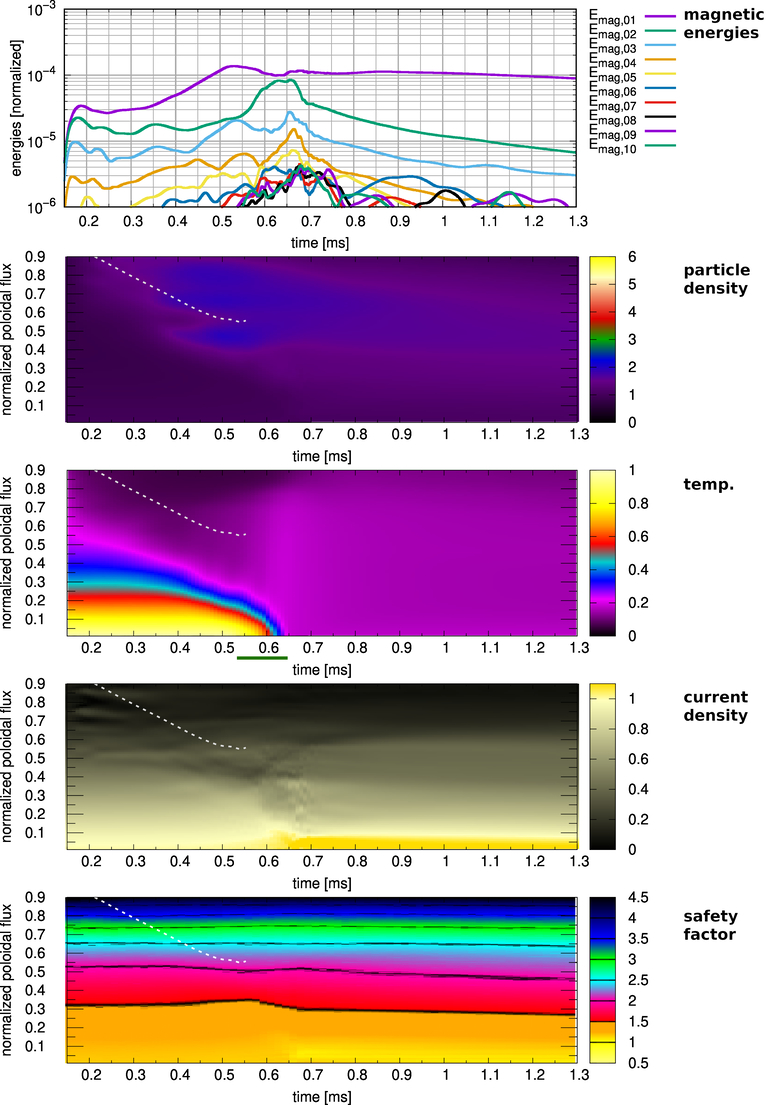}
  %\includegraphics[height=1.2\textwidth]{D-overview-TQ.jpg}
  %\includegraphics[height=1.2\textwidth]{overview-labels.jpg}
  % created with gnuplot by postproc/plot-density-vs-time-and-PsiN.gp
  % then combine-images.sh
\caption{Overview for case D ($10^{21}$ injected atoms, $30$ shards, $400\,\mathrm{m/s}$). From top to bottom, the plots show the time evolution of the magnetic energies, the density profile normalized to the initial core density, the temperature profile normalized to the initial core temperature, the current density profile normalized to the initial core current density and the profile of the safety factor q. The dashed line corresponds to the center of mass of the SPI shard cloud (until complete ablation).}
\label{fig:D-overview}
\end{figure}

\begin{figure}
\centering
  \includegraphics[width=0.6\textwidth]{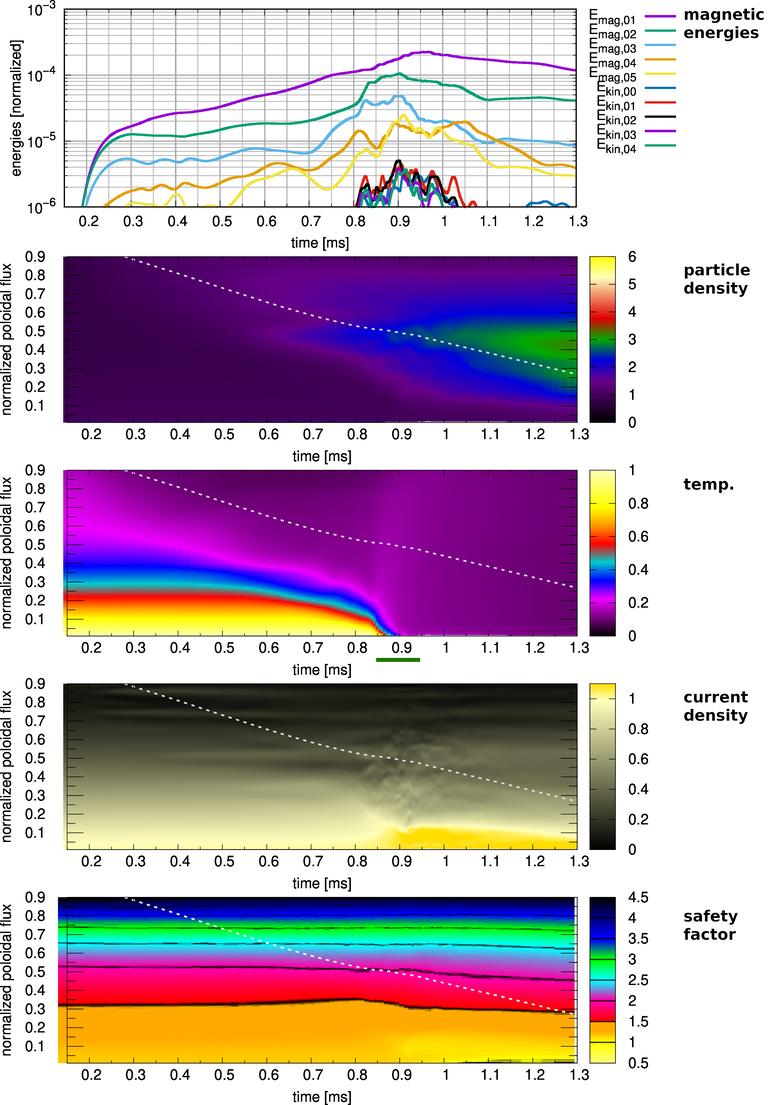}
  %\includegraphics[height=1.2\textwidth]{G-overview-TQ.jpg}
  %\includegraphics[height=1.2\textwidth]{overview-labels.jpg}
  % created with gnuplot by postproc/plot-density-vs-time-and-PsiN.gp
  % then combine-images.sh
\caption{Overview for case G ($3\cdot10^{21}$ injected atoms, single shard, $250\,\mathrm{m/s}$). From top to bottom, the plots show the time evolution of the magnetic energies, the density profile normalized to the initial core density, the temperature profile normalized to the initial core temperature \modif{(the green bar indicates the TQ time)}, the current density profile normalized to the initial core current density and the profile of the safety factor q. The dashed line corresponds to the center of mass of the SPI shard cloud.}
\label{fig:G-overview}
\end{figure}

\begin{figure}
\centering
  \includegraphics[width=0.6\textwidth]{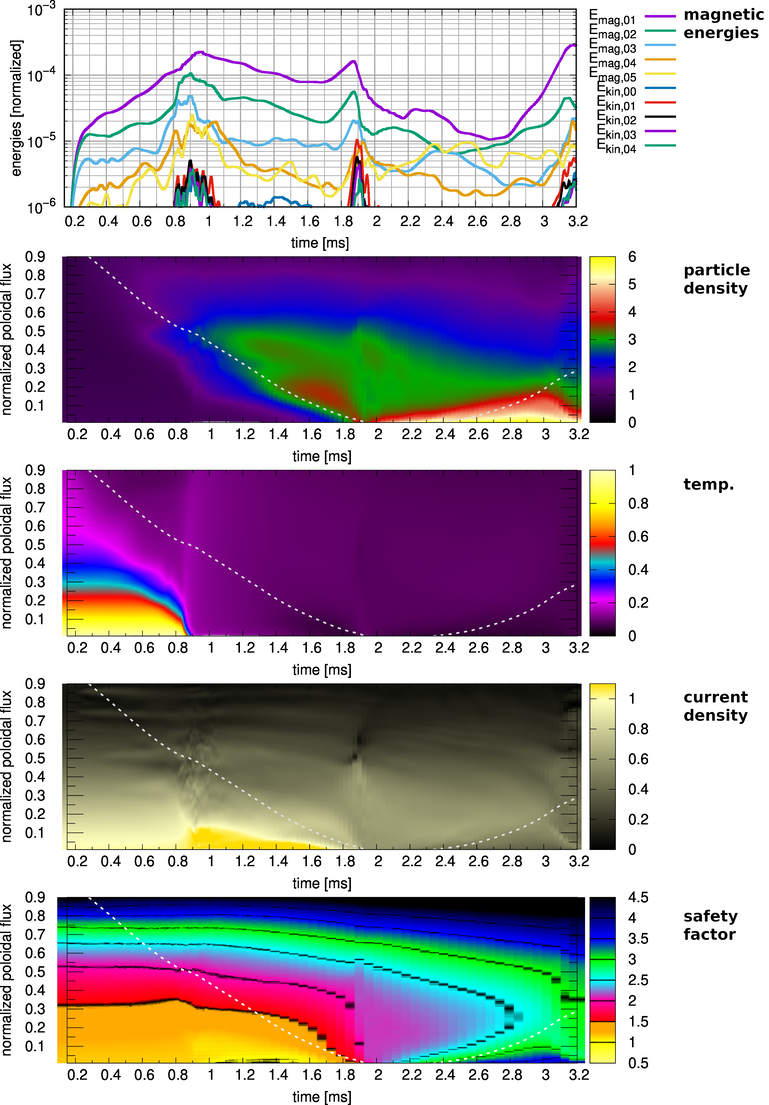}
  %\includegraphics[height=1.2\textwidth]{G-overview2.jpg}
  %\includegraphics[height=1.2\textwidth]{overview-labels.jpg}
  % created with gnuplot by postproc/plot-density-vs-time-and-PsiN.gp
  % then combine-images.sh
\caption{Overview for case G ($3\cdot10^{21}$ injected atoms, single shard, $250\,\mathrm{m/s}$) with an extended time range. From top to bottom, the plots show the time evolution of the magnetic energies, the density profile normalized to the initial core density, the temperature profile normalized to the initial core temperature, the current density profile normalized to the initial core current density and the profile of the safety factor q. The dashed line corresponds to the center of mass of the SPI shard cloud (pellet shards cross the center).}
\label{fig:G-overview2}
\end{figure}

\begin{figure}
\centering
  \includegraphics[width=0.6\textwidth]{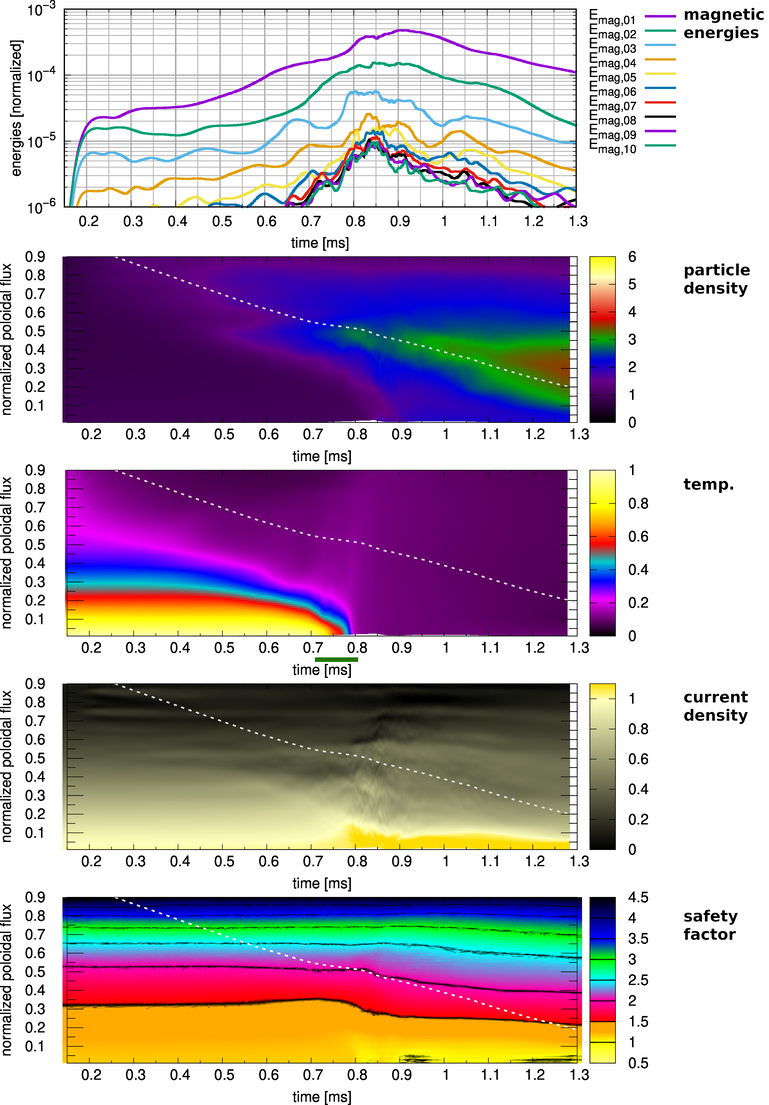}
  %\includegraphics[height=1.2\textwidth]{N-overview-TQ.jpg}
  %\includegraphics[height=1.2\textwidth]{overview-labels.jpg}
  % created with gnuplot by postproc/plot-density-vs-time-and-PsiN.gp
  % then combine-images.sh
\caption{Overview for case N ($3\cdot10^{21}$ injected atoms, $3$ shards, $250\,\mathrm{m/s}$). From top to bottom, the plots show the time evolution of the magnetic energies, the density profile normalized to the initial core density, the temperature profile normalized to the initial core temperature \modif{(the green bar indicates the TQ time)}, the current density profile normalized to the initial core current density and the profile of the safety factor q. The dashed line corresponds to the center of mass of the SPI shard cloud.}
\label{fig:N-overview}
\end{figure}

\begin{figure}
\centering
  \includegraphics[width=0.6\textwidth]{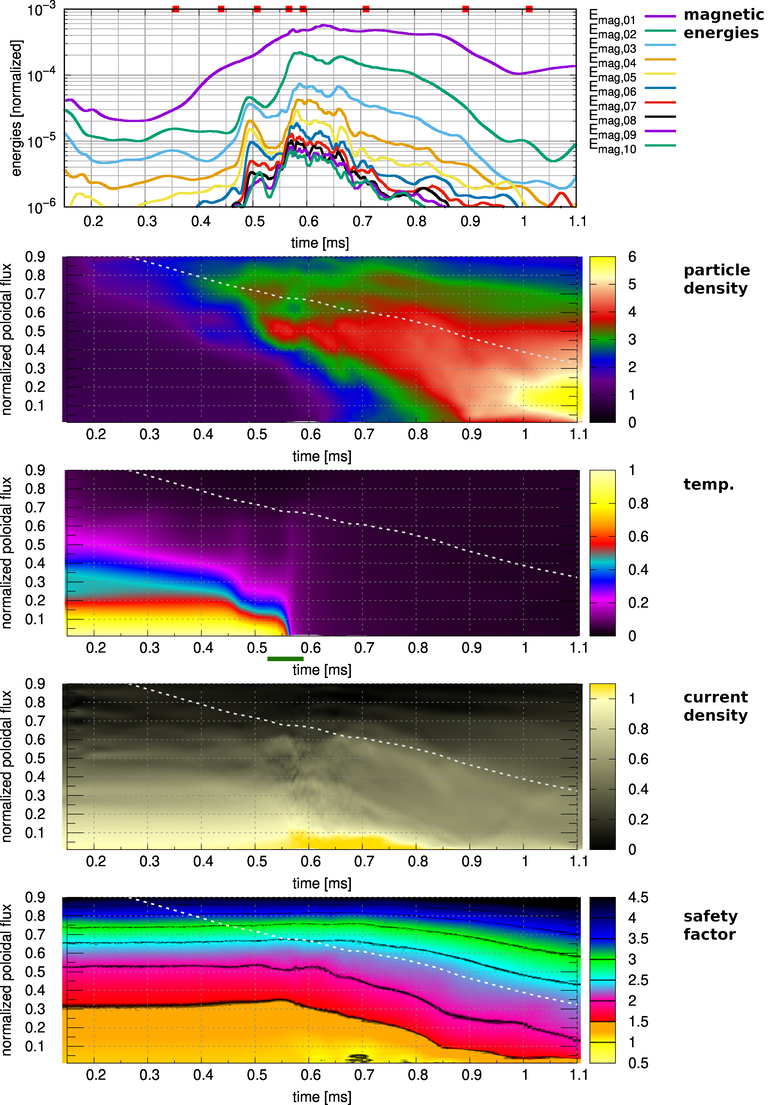}
  %\includegraphics[height=1.2\textwidth]{W-overview-TQ.jpg}
  %\includegraphics[height=1.2\textwidth]{overview-labels.jpg}
  % created with gnuplot by postproc/plot-density-vs-time-and-PsiN.gp
  % then combine-images.sh
\caption{Overview for case W ($10^{22}$ injected atoms, $100$ shards, $250\,\mathrm{m/s}$). From top to bottom, the plots show the time evolution of the magnetic energies, the density profile normalized to the initial core density, the temperature profile normalized to the initial core temperature \modif{(the green bar indicates the TQ time)}, the current density profile normalized to the initial core current density and the profile of the safety factor q. The dashed line corresponds to the center of mass of the SPI shard cloud. The red dots indicate time points analyzed in more detail in Section~\ref{:results:dynamics} and Figures~\ref{fig:poinc}--\ref{fig:W-convectioncells}.}
\label{fig:W-overview}
\end{figure}

In this section, the evolution of poloidally and toroidally averaged profiles is compared across several cases to highlight common properties and differences. \modif{Four representative cases are selected for this purpose, which cover the range from small to high amounts of injected material, and consequently from an incomplete TQ to a very violent TQ}. Figures~\ref{fig:D-overview} to~\ref{fig:W-overview} correspond to the cases D, G, N, and W, which show an increasingly violent TQ. In each figure, the uppermost plot shows the evolution of the magnetic perturbation energies versus time, the second plot shows the evolution of the density profile versus time, the third plot shows the evolution of the temperature profile versus time, the fourth plot shows the evolution of the current profile versus time, and the fifth plot shows the evolution of the q-profile versus time. The poloidally/and toroidally averaged quantities are plotted against the normalized poloidal flux (note that the definition of the normalized poloidal flux evolves slightly with time, since the magnetic field is changing). The white dotted line indicates the location of the center of mass of the SPI shard cloud \modif{calculated from the actual evolution of location and size of each shard}.

\paragraph{Case D} ($10^{21}$ injected atoms, $30$ shards, $400\,\mathrm{m/s}$).
As shown in Figure~\ref{fig:D-overview}, ablation approximately completes when the shard cloud reaches the q=2 surface. The core TQ takes place roughly at this time as well. An incomplete TQ is observed with a high post-TQ temperature and only a moderate density increase mostly outside the $q=2$ surface. The magnetic n=2 energy increases strongly when the core TQ takes place, while the n=1 magnetic energy stays roughly constant. The coupling to the higher harmonics is not strong. The current profile peaks a bit after the TQ, but not strongly enough for a $q=1$ surface to appear. This peaking of the current after the TQ appears since the edge is cooled first and the current is lost from the stochastic region and re-induced in the still hot plasma center like described in Ref.~\cite{Artola2019}. The q-profile is altogether only weakly perturbed. Following the core TQ, a high temperature ``plume'' is visible in the temperature profile, moving outwards across the whole plasma on a fast timescale by stochastic energy transport. This leads to a significant re-heating of the outer plasma region.

\paragraph{Case G} ($3\cdot10^{21}$ injected atoms, single shard, $250\,\mathrm{m/s}$).
As shown in from Figure~\ref{fig:G-overview}, the TQ approximately takes place when the shard cloud reaches the q=2 rational surface. Ablation is incomplete. The post TQ temperature is lower than the one of case D, but still relatively high indicating an incomplete TQ. Density still increases after the TQ, but mostly outside the location of the shard cloud. The magnetic n=2 and n=3 amplitudes increase strongly when the core TQ occurs, also n=1 is growing but not as sharply. \modif{The overall energy amplitudes are higher and a stronger coupling of the harmonics is observed than in case D.} The current profile peaks and a q=1 surface appears in the very center of the plasma. The plume of the outward temperature flow from the core TQ is again clearly visible and re-heats the plasma edge.

As shown in Figure~\ref{fig:G-overview2}, which shows the same data for the same case, but over a longer time window, strong bursts of MHD activity are seen again at about 1.9 ms and about 3.1 ms in the later evolution of case G. The burst at 1.9 ms appears roughly when q on axis drops below 1.5. After this crash, the q=1.5 and q=2 surfaces have disappeared and two q=2.5 surfaces are present due to a low axis current density. The q=2.5 surfaces finally merge and disappear around 2.8 ms, before the next burst appears around 3.1ms. This crash seems to occur, when one of the two q=3 surfaces reaches the plasma center. The plasma core density rises to very high values in this case, since ablation continues after the TQ due to the comparably high post-TQ temperature, and the \modif{strong radial mixing by the series of MHD bursts.}

\paragraph{Case N} ($3\cdot10^{21}$ injected atoms, $3$ shards, $250\,\mathrm{m/s}$).
As shown in Figure~\ref{fig:N-overview}, the TQ appears approximately when the shard cloud has reached the q=2 surface. The post TQ temperature is far lower than in the previous cases. Still the density continues to increase a bit after the TQ. During the TQ, the magnetic perturbation increases strongly across all harmonics due to strong mode coupling. The current profile peaks enough for a q=1 surface to appear shortly after the TQ in the very center of the plasma. The outward flowing heat is clearly visible in the temperature profile and shows a sub-structure which corresponds well to the evolution of the magnetic perturbation energies.

\paragraph{Case W} ($10^{22}$ injected atoms, $100$ shards, $250\,\mathrm{m/s}$).
As shown in Figure~\ref{fig:W-overview}, the TQ appears approximately when the shard cloud has reached the q=2.5 surface. A crash of the outer plasma core up to and including the q=1.5 surface is already observed when the shard cloud has reached the q=3 surface. The post TQ temperature is far lower in this case, however the density still increases slightly after the TQ even inside the location of the shard cloud due to a radial mixing effect. Both at the time of the crash of the outer plasma core and at the time of the core TQ, the magnetic energy of all harmonics rises in a tightly coupled way. An outflow of the thermal energy from the hot plasma region is seen already at $0.25\mathrm{ms}$ moving slowly in the radial direction due to the moderate stochastization at that point in time. Later on, at the two major crashes ($\approx0.46\mathrm{ms}$ and $\approx0.55\mathrm{ms}$), the radial propagation velocity of the outflow is significantly higher. These radial propagation velocities approximately are $1\,\mathrm{km/s}$, $10\,\mathrm{km/s}$, and $20\,\mathrm{km/s}$ reflecting the different stochastization levels at these points in time: \modif{The radial transport is approximately proportional to the magnetic perturbation energies consistently with theoretical expectations in a stochastic regime.} The current peaks after the TQ such that a q=1 surface is observed for a short while. Since a comparably fast current quench sets in due to the low post-TQ temperature, the overall q-profile starts to rise quickly after the TQ.

\modif{In the present Section, plasma dynamics were shown in detail for four representative simulations by analyzing the time evolution of radial profiles of density, temperature, current density and safety factor together with the magnetic perturbation amplitudes. The maximum density increases by a factor of six in the simulation with $1\cdot10^{22}$ atoms injected by 100 shards. The core density only increases about $200\,\mu\mathrm{s}$ after the core TQ. Whenever the stochastic region extends deeper into the plasma core, the outflow of thermal energy is visible by an increase of the temperature in the outer plasma regions. The radial propagation velocity of the thermal energy through the stochastic plasma is about $20\,\mathrm{km/s}$ at the time of strongest magnetic perturbations.}
In the following Section, case W \modif{showing the strongest increase in the plasma particle inventory} is analyzed in more detail.

%**************************************
\subsection{Evolution of density, temperature, plasma flows, and magnetic topology across the thermal quench}\label{:results:dynamics}
%**************************************

In this Section, case W is analyzed in detail which shows a pronounced TQ (see Figure~\ref{fig:W-overview}). In particular, the evolution of the density distribution, the temperature distribution, the plasma flows, and the magnetic topology are investigated. \modif{This section describes the dynamics observed in the simulation in detail allowing for qualitative comparisons to experiments when they become available. Since quantitative conclusions can not easily be drawn due to increased resistivity and a simplified shard cloud setup, only few quantitative conclusions are drawn.}

\begin{figure}
\centering
  \includegraphics[width=0.6\textwidth]{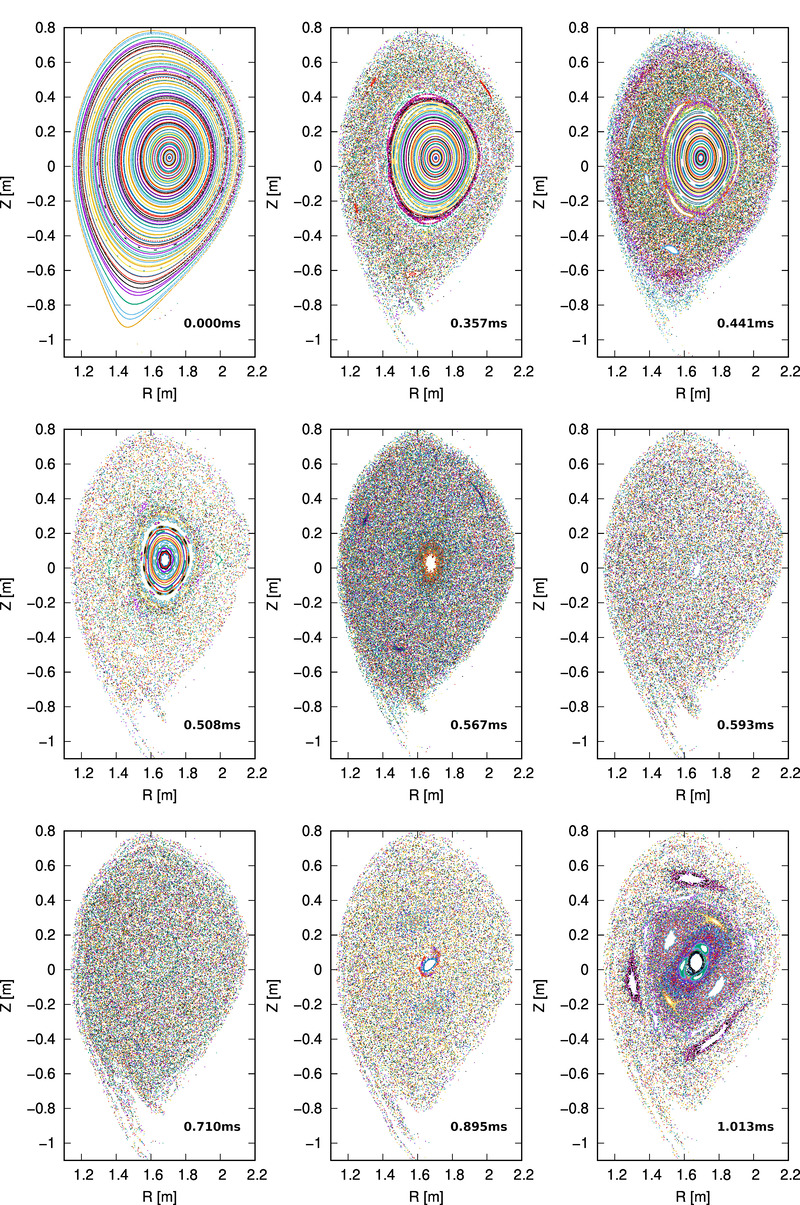}
  %\includegraphics[width=0.28\textwidth]{p0.jpg}
  %\includegraphics[width=0.28\textwidth]{p1.jpg}
  %\includegraphics[width=0.28\textwidth]{p2.jpg}
  %\includegraphics[width=0.28\textwidth]{p3.jpg}
  %\includegraphics[width=0.28\textwidth]{p4.jpg}
  %\includegraphics[width=0.28\textwidth]{p5.jpg}
  %\includegraphics[width=0.28\textwidth]{p6.jpg}
  %\includegraphics[width=0.28\textwidth]{p7.jpg}
  %\includegraphics[width=0.28\textwidth]{p8.jpg}
  % Scripts: poinc-for.sh and plot-all-poinc.sh
  % time points 0, 4000, 5000, 5800, 6500, 6800, 8200, 10400, 11800
\caption{The magnetic topology in case W is shown at several different points in time by Poincar\'e plots. These plots are created by starting 90 field lines across the whole plasma domain and tracing them for 1200 toroidal turns (or until they reach the divertor targets). Time points 0.000, 0.357, 0.441, 0.508, 0.567, 0.593, 0.710, 0.895, 1.013 are shown (all times in milliseconds).}
\label{fig:poinc}
\end{figure}

The magnetic topology before, during, and after the TQ is shown in Figure~\ref{fig:poinc} via Poincar\'e plots for several time points during the simulation. \modif{These time points are marked by red dots in Figure~\ref{fig:W-overview}.} The first plot represents the unperturbed equilibrium. The second plot corresponds to $t=0.357\,\mathrm{ms}$, where the whole edge region of the plasma is already stochastic up to the 3/2 magnetic surface. The next time point $t=0.441\,\mathrm{ms}$ has a significantly higher $n=1$ magnetic perturbation energy, while the higher harmonic perturbations have hardly changed. The stochastic region has not yet moved further inwards. At time point $t=0.508\,\mathrm{ms}$, the $n=2$ magnetic perturbation has strongly increased corresponding to a stochastization of the 3/2 surface. Time point $t=0.567\,\mathrm{ms}$ during the core TQ shows stochastization of the complete plasma, however weak stochastization at the grid center. During the following two time points $t=0.593\,\mathrm{ms}$ and $t=0.710\,\mathrm{ms}$ the complete plasma domain remains stochastic. Time point $t=0.895\,\mathrm{ms}$ corresponds to the onset of flux surface re-formation starting from the magnetic axis. The last time point shown at $t=1.013\,\mathrm{ms}$ is weakly stochastic in the whole plasma core such that runaway electrons (REs) would not be lost easily any more from this region, if they were present. Island remnants are visible at this time point at various rational surfaces in the stochastic region.

\begin{figure}
\centering
  \includegraphics[width=0.6\textwidth]{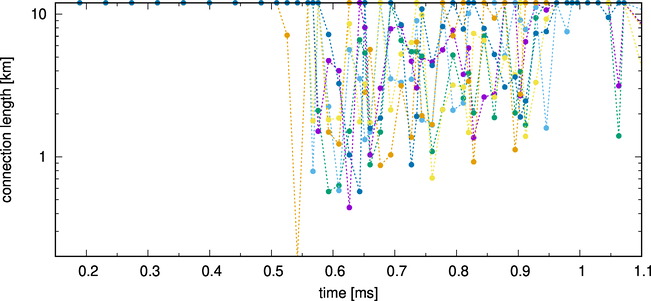}
  %\includegraphics[width=0.79\textwidth]{W-connection-length.jpg}
  % Scripts: poinc-for.sh like previous plot
  % then poinc-statistics.sh
  % then plot-poinc-stat-for-center.gp
\caption{Case W. For six field field lines started very close to the magnetic axis, the connection length to the divertor target is plotted versus time (maximum value of 12 km since the field line tracing is stopped then).}
\label{fig:poinc-stat}
\end{figure}

Figure~\ref{fig:poinc-stat} displays the connection length to the divertor of several field lines started close to the (initial) magnetic axis. Before the TQ, the field lines are never lost, during and after the TQ the connection length can drop to a few hundred meters corresponding to less than 100 toroidal turns. However, even in the phase of strong stochastization, time points exist, where some field lines remain confined inside the plasma. This motivates further analysis in the future regarding RE dynamics in such fields, to understand RE generation and losses during such a plasma evolution (such as the work for massive gas injection simulations for JET reported in Ref.~\cite{Sommariva2017,Sommariva2018}). \modif{The ratio of the parallel electric field to the critical electric field for RE avalanching increases from values around 0.3 before the TQ to values around 10 after the TQ in the plasma center for case W. These values were calculated for the realistic Spitzer resistivity, although an increase value had been used in the simulation. In spite of the value increasing beyond unity, the formation of a RE beam in this Deuterium injection case is very unlikely, since the electrical field \modifB{remains in the range of few percent of the Dreicer electrical field or below throughout most of the simulation time. Furthermore}, stochastization of the complete plasma lasting for several hundred micro-seconds \modifB{(including the times during which $E_{||}/E_\text{Dreicer}$ becomes largest}) would lead to a loss of the majority of the RE seed in case it was formed (dominated by the Dreicer mechanism in case of AUG).}

%In Figure~\ref{fig:W-dPsidt}, $d\Psi/dt$ is plotted for case $W$ at the same time points as for the previous figures. xxx

\begin{figure}
\centering
  \includegraphics[width=0.6\textwidth]{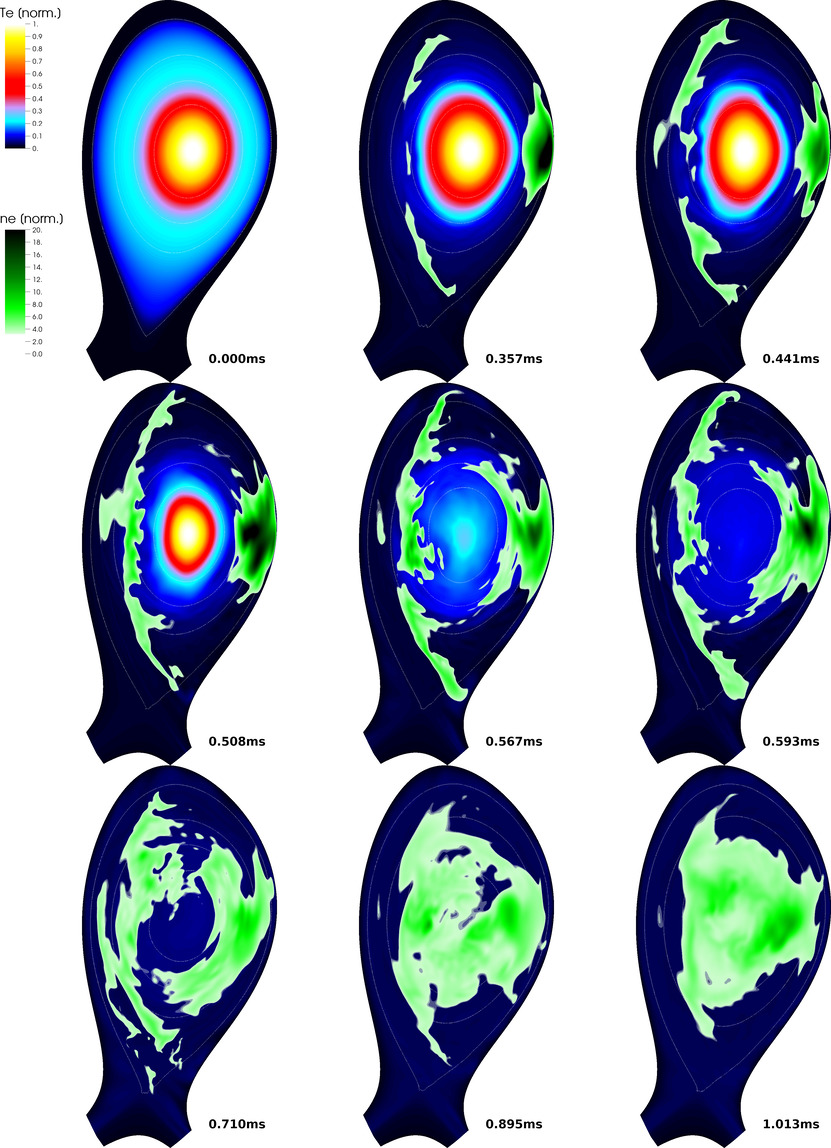}
  %\includegraphics[width=0.28\textwidth]{W-dens-temp-00000.png}
  %\includegraphics[width=0.28\textwidth]{W-dens-temp-04000.png}
  %\includegraphics[width=0.28\textwidth]{W-dens-temp-05000.png}
  %\includegraphics[width=0.28\textwidth]{W-dens-temp-05800.png}
  %\includegraphics[width=0.28\textwidth]{W-dens-temp-06500.png}
  %\includegraphics[width=0.28\textwidth]{W-dens-temp-06800.png}
  %\includegraphics[width=0.28\textwidth]{W-dens-temp-08200.png}
  %\includegraphics[width=0.28\textwidth]{W-dens-temp-10400.png}
  %\includegraphics[width=0.28\textwidth]{W-dens-temp-11800.png}
  % Scripts: poinc-for.sh and plot-all-poinc.sh
\caption{Density and temperature distributions are plotted for case W at the toroidal location $\phi=0$. Time points shown are the same as in Figure~\ref{fig:poinc}.}
\label{fig:denstemp}
\end{figure}

For the same time points as the Poincar\'e plots, Figure~\ref{fig:denstemp} shows the evolution of the density and temperature in a poloidal cross section. Strong 3D structures are present before the TQ. During the TQ and for $200\,\mu\mathrm{s}$ after it, the density profile remains hollow. About $400\,\mu\mathrm{s}$ after the TQ, the core density has increased by a factor of about six compared to the initial equilibrium state. Favorably for RE mitigation/suppression, the re-formation of flux surfaces only appears, when the density profile is not hollow any more.

\begin{figure}
\centering
  \includegraphics[width=0.6\textwidth]{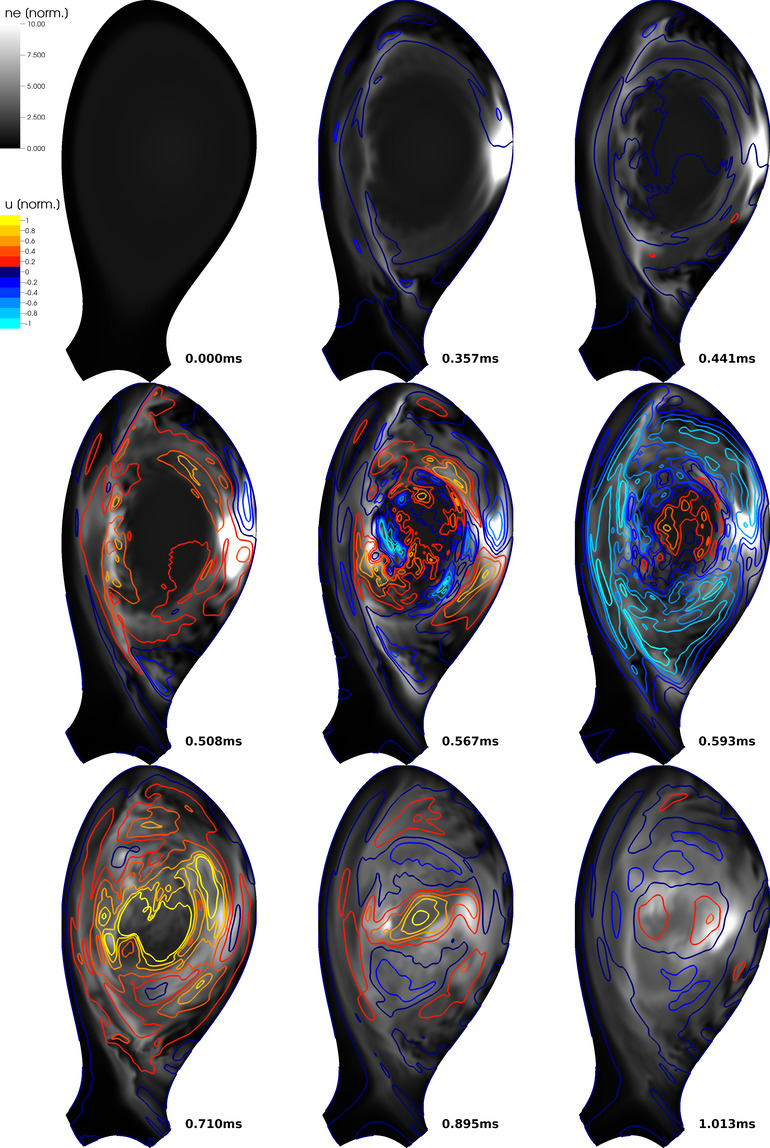}
  %\includegraphics[width=0.28\textwidth]{W-streamfunc-00000.png}
  %\includegraphics[width=0.28\textwidth]{W-streamfunc-04000.png}
  %\includegraphics[width=0.28\textwidth]{W-streamfunc-05000.png}
  %\includegraphics[width=0.28\textwidth]{W-streamfunc-05800.png}
  %\includegraphics[width=0.28\textwidth]{W-streamfunc-06500.png}
  %\includegraphics[width=0.28\textwidth]{W-streamfunc-06800.png}
  %\includegraphics[width=0.28\textwidth]{W-streamfunc-08200.png}
  %\includegraphics[width=0.28\textwidth]{W-streamfunc-10400.png}
  %\includegraphics[width=0.28\textwidth]{W-streamfunc-11800.png}
  % visit0000.session
\caption{For case W, the density distribution is shown in black and white. Contours of $u$, the stream function of the perpendicular plasma velocity, are plotted in color. Time points shown are the same as in Figure~\ref{fig:poinc} and~\ref{fig:denstemp}. An outwards oriented convection cell is observed in the high density region whenever heat is released from the plasma core such that the pressure of the high density cloud is suddenly increased. This is particularly visible at 0.567ms, which corresponds to the core TQ time. After the TQ, strong poloidal rotation induced by Maxwell stress is visible (t=0.710ms), which supports the re-healing of flux surfaces by decoupling MHD modes at different radial locations.}
\label{fig:W-convectioncells}
\end{figure}

Figure~\ref{fig:W-convectioncells} shows contours of the stream function $u$ of the poloidal velocity for the same time points as those of Figures~\ref{fig:poinc} and~\ref{fig:denstemp}. Strong outwards oriented convection cells form at the location of the high density cloud several times during the simulation (a gradient of $u$ pointing downwards corresponds to a convection in $+\mathbf{e}_R$ direction), hindering the penetration of the ablated material into the plasma core. These cells always form, when a significant amount of thermal energy is released from the hot plasma core, either by the core TQ itself, or by a loss of a layer of closed flux surfaces (inward motion of the stochastic front). It is observed particularly strongly between $0.47\,\mathrm{ms}$ and $0.59\,\mathrm{ms}$. The convection cells are visible for instance at the time points $0.508\,\mathrm{ms}$ and $0.567\,\mathrm{ms}$ in Figure~\ref{fig:W-convectioncells} and are related to the plasmoid drifts described in Ref~\cite{Mueller1999}, which render material assimilation for LFS injection inefficient. For HFS injection, the corresponding interchange instabilities would lead to a material convection towards the magnetic axis instead, thus greatly increasing the material assimilation at this stage.

As seen from time points $0.567\,\mathrm{ms}$ and $0.593\,\mathrm{ms}$, strongly localized convection cells form in the plasma center during the TQ. However, these cannot increase the core plasma density immediately, since they are spatially separated from the high density cloud, which had been pushed outwards by the interchange instability driven convection cell described above. After the TQ, strong poloidal rotation resulting from Maxwell stress is observed as seen from time point $0.710\,\mathrm{ms}$. In addition, convection cells oriented towards the plasma center are now forming as seen from time points $0.710\,\mathrm{ms}$ and $0.895\,\mathrm{ms}$, which lead to an increase of the core plasma density turning the hollow density profile again into a monotonic one. In addition, parallel transport with the ion sound speed in the stochastic field contributes to an increase of the core density. Together, these parallel and perpendicular transport mechanisms fill the hole in the density profile within about $0.2\,\mathrm{ms}$ after the TQ. About $0.4\,\mathrm{ms}$ after the TQ, the density distribution has become smooth and monotonic across the plasma and plasma flows have dropped to low values.

\modif{An SPI simulation leading to a particularly strong increase in the plasma particle inventory (case W) was analyzed in detail in this section. The magnetic field is fully stochastic during the TQ and for several $100\,\mu\mathrm{s}$ afterwards. While the core density has not increased significantly at the time of the TQ, it rises considerably before the re-formation of flux surfaces sets in starting from the magnetic axis, which is a favorable observation for RE suppression/mitigation. Radially outwards oriented convection cells form whenever thermal energy is released from the hot core of the plasma and heat the high density cloud around the ablation region. These convection cells are responsible for inefficient material assimilation in case of low field side injection.}

%**************************************
\section{Conclusions and Outlook}\label{:conclusions}
%**************************************

First predictive simulations for deuterium shattered pellet injection into an ASDEX Upgrade H-mode plasma are shown. Various simulations were performed to study the influence of plasma parameters and injection parameters onto the dynamics. In spite of some limitations of the present simulations (e.g., deuterium injection only, neglected impurity background radiation, simplified modelling of the shard cloud, artificially increased resistivity in most cases), various insights into the MHD activity and TQ dynamics during SPI is obtained. The number of atoms injected, the number of shards, the injection velocity, the plasma resistivity, and the parallel heat conductivity influence the dynamics strongly. \modif{An increase of each of these parameters generally is seen to act destabilizing. In tendency, the TQ occurs earlier and leads to a lower post-TQ temperature. The TQ duration strongly decreases for higher plasma resistivity, while the dependency on other parameters is less clear.}

\modif{One aim of the article is to show how MHD activity affects shard ablation and density increase.} The influence of MHD activity onto the ablation of shards was analyzed by comparing the self-consistent simulation with simplified simulations without MHD activity and a fixed background profile. The MHD activity strongly influences ablation, in particular stochastic heat transport from the originally confined region leads to a fast loss of thermal energy reducing ablation.

\modif{We show which characteristic features are observed during the TQ.} A loss of the plasma edge confinement is observed almost instantaneously when the shards reach the plasma boundary, since an H-mode plasma is considered. The loss of edge plasma current and the corresponding steepening of the current profile are shown to affect the destabilization of MHD activity significantly. \modif{Also the helical perturbation of the temperature causing a helical perturbation of the resistivity and consequently a helical perturbation of the current density is shown to affect mode destabilization considerably.} A thermal quench (TQ) is observed within the first millisecond after injection in most simulations and takes typically about 0.1 milliseconds. Compared to previous simulations for JET~\cite{Hu2018}, the stochastic front is moving inwards a lot faster. This may be related to a different ordering between the shard propagation time scale and the time scale the current profile takes to adapt (the injection velocity is slower in the considered case, while the resistivity is higher). At realistic plasma resistivity \modif{for which only a single simulation has been performed in the present study due to the higher computational costs}, the thermal quench appears later after injection and takes significantly longer.

\modif{With this article, we aim to clarify how plasma and SPI parameters affect the TQ dynamics, how the density profile evolves during the TQ, and how the core density increases before flux surface reformation after the TQ.} At $10^{21}$ atoms injected, only incomplete TQs are observed, at $3\cdot10^{21}$ the temperature collapses are more complete, and at $1\cdot10^{22}$ atoms they are complete (low post-TQ temperature) and more violent. The shards have not crossed the q=2 rational surface before the TQ onset in almost all cases. The stochastic outflow of thermal energy from the plasma core leads to the formation of a convection cell in the high density cloud oriented radially outwards for the low field side injection considered here, hindering an efficient assimilation and penetration of the material into the core. The density profile has a strong 3D structure before and during the TQ. It remains hollow for about 0.2 milliseconds after the TQ. Successive mixing by small convection cells, \modif{parallel transport in the stochastic field, and} inward transport by a large convection cell leads to a flat density profile later on. With $10^{22}$ atoms injected, an increase of the initial plasma core density by a factor of about six is observed. At the onset of the TQ, the magnetic topology becomes fully stochastic and the connection length of field lines traced from the region of the magnetic axis to the divertor targets drops to few hundred meters. First flux surfaces in the core as well as island remnant surfaces begin to re-form a few hundred microseconds after the onset of the TQ. This re-formation of flux surfaces is enforced also by strong poloidal rotation driven by Maxwell stress, observed in particular after the TQ. The re-formation of flux surfaces sets in at a point in time, when the density profile is not hollow any more, which is a promising observation for RE mitigation or suppression.

\modif{We try to give an estimate for the amount of material needed to trigger a TQ.} Taking into account material losses before the shards have reached the plasma edge, a less violent plasma response at fully realistic resistivity (which was used only in one simulation for computational reasons), and the stabilizing effect of plasma flows prior to the injection, an initial pellet size of around $(8\pm4)\cdot10^{21}$ deuterium atoms is expected to cause a TQ in the experiment promptly after injection. A delayed TQ several milliseconds after injection would likely occur already with significantly smaller amounts of injected material (density limit disruption).

Future work will take into account impurity background radiation, plasma background flows, MHD activity like NTMs possibly existing before the injection, fully realistic plasma parameters, and will be adapted to the results of a lab characterization of the shard cloud in ASDEX Upgrade. For impurity SPI, an extended impurity model will be implemented in JOREK \modif{(conclusions for impurity SPI based on our present simulations of deuterium SPI are not directly possible)}. Using the free boundary JOREK-STARWALL code~\cite{Hoelzl2012B}, also vertical stability and control will be taken into account and runaway electron dynamics will be studied using the models described in Refs.~\cite{Sommariva2017,Bandaru2019A}. Numerical improvements which are presently ongoing, are expected to reduce the computational costs and enhance numerical stability at fully realistic parameters. When the ASDEX Upgrade SPI injector becomes available, detailed validations against experiments are planned.

%**************************************
\section{Acknowledgements}\label{:ack}
%**************************************

Parts of this work has been carried out within the framework of the EUROfusion Consortium and has received funding from the Euratom research and training programme 2014-2018 and 2019-2020 under grant agreement No 633053. Parts of this work was carried out using the Marconi-Fusion supercomputer. The views and opinions expressed herein do not necessarily reflect those of the European Commission. The authors would like to acknowledge many fruitful discussions and the efficient code development in the JOREK team. Helpful discussions with Gergely Papp, Gabriella Pautasso, Vinodh Kumar Bandaru, Karl Lackner and Sibylle G\"unter are acknowledged in particular.

%**************************************
\bibliography{mybib}
%**************************************

%\appendix
%
%\section{A few parameters (only for myself)}
%
%\begin{align}
%n_{e,0} &= 7.5\cdot10^{19}m^{-3} \\
%m_\text{ion} &= m_\text{deuteron} \\
%\rho_0 &= 2.5\cdot10^{-7}kg\,m^{-3} \\
%\sqrt{\mu_0/\rho_0} &= 2.23 \\
%\sqrt{\mu_0 \rho_0} &= 5.6\cdot10^{-7} \\
%\end{align}

% -----------------
% -----------------

% -----------------
% -----------------

% -----------------
% -----------------

% -----------------
% -----------------

% -----------------
% -----------------

% -----------------
% -----------------

% -----------------
% -----------------

% -----------------
% -----------------

% -----------------
% -----------------

% -----------------
% -----------------

% -----------------
% -----------------

% -----------------
% -----------------

% -----------------
% -----------------

% -----------------
%\begin{figure}
%\centering
%  \includegraphics[width=0.28\textwidth]{W-dPsidt-00000.png}
%  \includegraphics[width=0.28\textwidth]{W-dPsidt-04000.png}
%  \includegraphics[width=0.28\textwidth]{W-dPsidt-05000.png}
%  \includegraphics[width=0.28\textwidth]{W-dPsidt-05800.png}
%  \includegraphics[width=0.28\textwidth]{W-dPsidt-06500.png}
%  \includegraphics[width=0.28\textwidth]{W-dPsidt-06800.png}
%  \includegraphics[width=0.28\textwidth]{W-dPsidt-08200.png}
%  \includegraphics[width=0.28\textwidth]{W-dPsidt-10400.png}
%  \includegraphics[width=0.28\textwidth]{W-dPsidt-11800.png}
%  % visit0000.session
%\caption{Case W. The spatial distribution of $d\Psi/dt$ is plotted for the same time points as the Poincar\'e plots.}
%\label{fig:W-dPsidt}
%\end{figure}
% -----------------

% -----------------
% -----------------

% -----------------
% -----------------

\end{document}